\documentclass[aps,showpacs,preprint,11pt]{revtex4}

\usepackage{amsmath,amssymb}
\usepackage[dvips]{graphics}
\usepackage{epsfig}
\usepackage{graphicx}
\usepackage{bm}
\usepackage{epsfig}
\usepackage{amsmath}

\newcommand{\beq}{\begin{equation}}
\newcommand{\eeq}{\end{equation}}
\newcommand{\beqa}{\begin{eqnarray}}
\newcommand{\eeqa}{\end{eqnarray}}
\newcommand{\beqar}{\begin{eqnarray*}}
\newcommand{\eeqar}{\end{eqnarray*}}

\newcommand\munu{\ensuremath{{\mu\nu}}}
\newcommand\pprime{\ensuremath{{\prime\prime}}}
\begin{document}

\title{Bonnor stars in d spacetime dimensions}

\author{Jos\'e P. S. Lemos}
\affiliation{Centro Multidisciplinar de Astrof\'{\i}sica -- CENTRA,
\\Departamento de F\'{\i}sica,
Instituto Superior T\'ecnico - IST,
Universidade T\'ecnica de Lisboa - UTL,
Av. Rovisco Pais 1, 1049-001 Lisboa, Portugal,
email: lemos@fisica.ist.utl.pt}
\author{Vilson T. Zanchin }
\affiliation{Centro de Ci\^encias Naturais e Humanas, Universidade Federal
do ABC, Rua Catequese 242, 09090-400, Santo Andr\'e, SP, Brazil,
email: zanchin@ufabc.edu.br}

\begin{abstract}

Bonnor stars are regular static compact configurations in equilibrium,
composed of an extremal dust fluid, i.e., a charged dust fluid where
the mass density is equal to the charge density in appropriate units
and up to a sign, joined to a suitable exterior vacuum solution, both
within Newtonian gravity and general relativity. In four dimensions,
these configurations obey the Majumdar-Papapetrou system of equations,
in one case, the system is a particular setup of Newtonian gravity
coupled to Coulomb electricity and electrically charged matter or
fluid, in the other case, the system is a particular setup of general
relativity coupled to Maxwell electromagnetism and electrically
charged matter or fluid, where the corresponding gravitational
potential is a specially simple function of the electric potential
field and the fluid, when there is one, is made of extremal dust.
Since the Majumdar-Papapetrou system can be generalized to $d$
spacetime dimensions, as has been previously done, and higher
dimensional scenarios can be important in gravitational physics, it is
natural to study this type of Bonnor solutions in higher dimensions,
$d\geq4$.  As a preparation, we analyze Newton-Coulomb theory with an
electrically charged fluid in a Majumdar-Papapetrou context, in
$d=n+1$ spacetime dimensions, with $n$ being the number of spatial
dimensions.  We show that within the Newtonian theory, in vacuum, the
Majumdar-Papapetrou relation for the gravitational potential in terms
of the electric potential, and its related Weyl relation, are
equivalent, in contrast with general relativity where they are
distinct. We study a class of spherically symmetric Bonnor stars
within this theory.  Under sufficient compactification they form point
mass charged Newtonian singularities. We then study the analogue type
systems in the Einstein-Maxwell theory with an electrically charged
fluid.  Drawing on our previous work on the $d$-dimensional
Majumdar-Papapetrou system, we restate some properties of this
system. We obtain spherically symmetric Bonnor star solutions in
$d=n+1$ spacetime dimensions. We show that these stars, under
sufficient compactification, form $d$-dimensional quasi-black
holes. We also show that in the appropriate low gravity limit theses
solutions turn into the solutions of Newtonian gravity, i.e., they are
quasi-Newtonian Bonnor stars. In this connection, we note that the
star solutions in Majumdar-Papapetrou Newtonian gravity, when
contrasted to those solutions in Majumdar-Papapetrou general
relativity, display clearly the branching off of the high density
objects that may arise in the strong field regime of each theory, mild
singularities in one theory, quasi-black holes in the other.  Another
important feature worth of mention is that, whereas there are no
solutions for Newtonian or relativistic stars supported by degenerate
pressure in higher dimensions, higher dimensional Bonnor stars,
supported by electric repulsion, do indeed have solutions within
Newtonian gravity and general relativity.  So the existence of stars
in higher dimensions depends on the number of dimensions itself, and
on the underlying field content of those stars.


\end{abstract}


\maketitle

\section {Introduction}
\label{introd}

\subsection{Definition}

Extremal charged dust, or simply extremal dust, is understood as
charged dust fluid, or matter, with the mass density being equal to
the charge density, in appropriate units, which implies that for each
such a dust particle, eventually composing a system, the gravitational
attraction is precisely balanced by the electric repulsion, both
within Newtonian gravity coupled to Coulomb electricity and an
electrically charged fluid or matter,  i.e., the Newton-Coulomb
with charged fluid system, and within general relativity coupled to
Maxwell electromagnetism and an electrically charged fluid, i.e.,
the Einstein-Maxwell with charged fluid system.  Bonnor stars are
then defined as regular static equilibrium configurations, in
Newtonian and general relativity contexts, composed of extremal dust,
with a finite boundary appropriately attached to an asymptotically
flat regular extremal charged vacuum, and where the configuration of
the matter dust can have any shape, a spherical symmetric shape being
of special interest, due to the added symmetry and due to the fact
that it can be joined to an asymptotically flat regular extremal outer
Reissner-Nordstr\"om spacetime in the general relativistic case, with
mass $M$ equal to charge $Q$ in appropriate units.  Bonnor stars
appear in $d=n+1$ spacetime dimensions, where $n$ is the number of the
spatial dimensions. The initial studies where performed for $d=4$.
Bonnor stars have also been called Majumdar-Papapetrou stars, but
here we reserve the name Majumdar-Papapetrou for the type of matter,
and name the whole system, namely Majumdar-Papapetrou matter plus
vacuum plus junction, as a Bonnor star.

\subsection{Four dimensional analyses}

\subsubsection{Context}
Such stars were studied mainly within general relativity,
although with some incursions onto Newtonian gravity,
by Bonnor \cite{bonnor53}-\cite{bonnor99} and in several other works,
see e.g., \cite{kleberlemoszanchin,lemoszanchin1} for Bonnor stars
properly said, and in \cite{lemosweinberg} for a variant where there
is no need for a junction.  One striking
property of these solutions found in
\cite{kleberlemoszanchin,lemoszanchin1,lemosweinberg} is that when
they approach their gravitational radius in a static sequence of
configurations, these stars do not form black holes, but rather
quasi-black holes, where a quasi-black hole is an object indistinguishable
to the exterior from a black hole but with different intrinsic
properties.  In \cite{lemoszaslavskii1,lemoszaslavskii2}
the properties of quasi-black
holes formed from Bonnor stars were studied in detail.
See also \cite{horvat} for a further study on the properties
of Bonnor stars.

\subsubsection{Vacuum Majumdar-Papapetrou solutions}

Within Newtonian gravity coupled to Coulomb electricity, i.e.,
the Newton-Coulomb system, electrically charged solutions in vacuum
represent charged point masses.  Within general relativity,
electrically charged solutions, in vacuum, have to be analyzed through
the Einstein-Maxwell system of equations, where one couples Einstein
gravity to Maxwell electromagnetism.  These solutions were found just
after general relativity was formulated. On one hand, Reissner
\cite{reissner}, then Nordstr\"om \cite{nordstrom}, then Jeffery
\cite{jeffery}, hit on the static vacuum charged spherically symmetric
solution, the Reissner-Nordstr\"om solution, with its two parameters,
the mass $M$ and the charge $Q$. We now know, that for ${\sqrt
G}M<\epsilon Q$ (we put the speed of light $c=1$ throughout), where
$G$ is Newton's gravitational constant in four spacetime dimensions
$(G_4\equiv G)$ and $\epsilon=1$ if the charge is positive and
$\epsilon=-1$ if the charge is negative, one has a charged naked
singularity, for ${\sqrt G}M>\epsilon Q$ one has a
Reissner-Nordstr\"om black hole, and ${\sqrt G}M=\epsilon Q$ one has
an extremal black hole (see \cite{eddingtonbook} for an early
discussion, and \cite{mtw} for a full discussion of this type of
solutions).

On the other hand, with the purpose of seeking vacuum
static solutions electrically charged, a different route was
originated from Weyl \cite{weyl}, the route that will take us to the
Bonnor stars \cite{bonnor53}-\cite{horvat}.  Wanting to go a
step further from spherical symmetry he sought axial symmetry.
Define the metric component $g_{00}$ as $g_{00}\equiv W^{2}$, where,
depending on the situation, it can be more convenient to define
$U\equiv W^{-1}$, i.e., $g_{00}\equiv U^{-2}$.  Then, Weyl asked himself,
within Einstein-Maxwell theory, what would happen if $W^{2}$, in a
static axisymmetric vacuum electric system, is to have a functional
dependence on the electric potential field $\varphi$, i.e., the Weyl
ansatz $g_{00}=g_{00}(\varphi)$ or equivalently $W=W(\varphi)$. He found
first what is now called the Weyl relation, i.e.,
$W^{2}=\left(a_0-\epsilon\sqrt{G}\,\varphi \right)^2 + b_0$,
where $a_0$ and $b_0$ are constants of
integration and $G$ is Newton's gravitational constant, and second
that the spatial components of the metric had to obey other specific
differential equations.  Majumdar \cite{majumdar} made several
improvements on Weyl's work. He showed that the Weyl relation, if it
existed was independent of the symmetry, axial or otherwise. But
further, he showed, still in vacuum, that if the relation was to be a
perfect square, so that, $W=a_0 -\epsilon\sqrt{G\,}\varphi$, then
the spatial part of the metric could be put in a simple form, as
$1/W^2$, i.e. $U^2$, times the flat spatial metric, and the
Einstein-Maxwell system of equations would reduce to one single
equation for $W$, i.e. for $U$, a Laplace equation in flat space. In
this perfect square case, one can show that specializing to spherical
symmetry, the mass $M$ of the solution is equal to its charge $Q$,
$\sqrt{G}M=\epsilon Q$.  This makes contact with the
Reissner-Nordstr\"om family of solutions through the extremal
solution, $\sqrt{G}M=\epsilon Q$, although not through the other ones,
since the Reissner-Nordstr\"om family generically does not admit a
functional relationship between the metric and the electric
potentials. These vacuum $\sqrt{G}M=\epsilon Q$ solutions were further
commented by Papapetrou \cite{papapetrou}, who also understood that
since the gravitational attraction is equal to the electric repulsion
for such objects one could have many discrete such objects scattered
at will in space, that it would also give a vacuum static
configuration solution, with no symmetry whatsoever. The perfect
square relation is usually called Majumdar-Papapetrou relation, as we
do here, although sometimes it is called, perhaps more appropriately
Weyl-Majumdar relation. The complete understanding of a single
extremal Reissner-Nordstr\"om, also a Majumdar-Papapetrou solution,
solution was achieved by Carter \cite{carter66}, through a
Carter-Penrose diagram, and the complete understanding of the vacuum
Majumdar-Papapetrou solutions, with many extremal black holes
scattered around was performed by Hartle and Hawking \cite{hartle},
who have done the maximal analytical continuation in the molds of
Carter \cite{carter66}.

\subsubsection{Beyond vacuum: dust Majumdar-Papapetrou solutions}

Things become more interesting if one goes beyond vacuum and puts
matter into the Newton-Coulomb system of equations and into the
Einstein-Maxwell system of equations.  Majumdar \cite{majumdar} and
Papapetrou \cite{papapetrou} understood this, and within general
relativity, showed, for some special restrictions on the metric
inspired from the vacuum case, such as the relation
$W=a_0-\epsilon\sqrt{G\,}\varphi$ (which in this case can be
considered an ansatz), one could find the system of equations yield a
single equation that moreover reduces to a Poisson equation, and in
which the mass density $\rho_{\rm m}$ times $\sqrt G$ is equal to the
charge density $\rho_{\rm e}$, up to a sign, ${\sqrt G}\rho_{\rm
m}=\epsilon\rho_{\rm e}$, with again $\epsilon=\pm1$.  That is, the
matter is made of an extremal dust fluid.  This is the
Majumdar-Papapetrou condition. Note that the relation between the
potentials we call Majumdar-Papapetrou relation, whereas the relation
between the densities we call Majumdar-Papapetrou condition.  As seen
in \cite{majumdar,papapetrou}, it is remarkable that a simple obvious
fact in Newton-Coulomb theory with an electrically charged fluid, that
if the mass density and charge density are equal (in appropriate
(geometric) units where $G=1$) then there is exact balancing of the
gravitational and electric forces throughout the matter and so there
is a static solution, also holds in Einstein-Maxwell theory with an
electrically charged fluid, with no need for further stresses, such as
pressure or tension. The basic feature of Majumdar-Papapetrou systems
is that they describe static spacetimes filled either with extremal
charged vacuum or extremal charged dust fluids, such that the metric
and electromagnetic fields may be characterized by two scalar
functions, namely, the redshift metric function $W$, i.e.  $U^{-1}$,
which plays the role of the gravitational potential, and the electric
potential $\varphi$, which in turn obey the Majumdar-Papapetrou
relation, $W=a_0 -\epsilon\sqrt{G\,}\varphi$. Further interesting
developments were achieved by Das \cite{das62}, and De and
Raychaudhuri \cite{de-raichaudhuri}, who considered charged dust
distributions in equilibrium, the way envisioned by Majumdar and
Papapetrou, and showed some other conditions related to the functional
form of the metric in terms of the electric potential and the equality
between mass and charge densities. Das \cite{das62} revealed that the
equality between the densities implies the functional form on the
potentials, and De and Raychaudhuri \cite{de-raichaudhuri} proved that
given the functional form above, and provided there are no
singularities in the distribution, the equality of mass and charge
densities follows directly from the field equations. There are other
interesting properties of Majumdar-Papapetrou systems in the context
of conformal static charged solutions \cite{lindenbellbicakkatz}.

\subsubsection{Bonnor stars: junction of dust with vacuum
Majumdar-Papapetrou solutions}

When one joins smoothly, within Newtonian gravity as well as
within general relativity, Majumdar-Papapetrou interior matter
solutions to Majumdar-Papapetrou exterior vacuum solutions, i.e., the
two types of solutions mentioned in the previous paragraphs, one
obtains Bonnor stars \cite{bonnor53}-\cite{bonnor99} and their
developments \cite{kleberlemoszanchin}-\cite{horvat}.
Bonnor stars could instead be called Majumdar-Papapetrou stars as was
done in \cite{lemoszanchin1}, but it is more proper to characterize
the matter part as a Majumdar-Papapetrou system, and this combined
with placing a boundary and the corresponding junction to a vacuum,
bringing together a whole lot of new properties, as a Bonnor
star. Throughout his works, Bonnor gradually improved the
understanding of the properties of these stars.

In the first two papers \cite{bonnor53,bonnor54} Bonnor worked out
aspects of electric Majumdar-Papape\-trou solutions in an axial
symmetric vacuum and extended these results through dualities to
magnetic solutions.  In \cite{bonnor60} a pre-Bonnor star is
developed, and it is noted that for ${\sqrt G}\rho_{\rm m}=\epsilon
\rho_{e}$, in a Majumdar-Papapetrou system, the gravitational mass of
the system is equal to the matter mass because the negative
gravitational self-energy of the distribution is balanced by the
corresponding positive electrical self-energy, also pointing out that
$\sqrt{G}M=\epsilon Q$ models can have various interests and
applications.  In \cite{bonnor64,bonnor65} Bonnor understood for the
first time that although delicate, the balance can exist, an atom
stripped off of an electron immersed in about $10^{18}$ atoms is
enough, and that the charge density can play an important part in the
equilibrium of large bodies, further suggesting that it may halt
gravitational collapse, at a time where large bodies studies were in
vogue due to the appearance of quasars. It is also mentioned that
bodies of arbitrary shape composed of such extremal dust can exist,
constructing explicitly a spherically symmetric solution, the first
Bonnor star. The way it is constructed delineates a standard way of
finding such type of solutions.  Assuming a given form for the
gravitational potential $U$ one can find the density distribution, and
one hopes that that assumption yields a physical distribution of
charged dust matter. It is not a method for solving the differential
equation of the Majumdar-Papapetrou problem, it is an art of correct
guessing.  In \cite{bonnor72,bonnor75}, both works in collaboration
with Wickramasuriya, interesting physical properties of some Bonnor
stars are discussed.  In particular in \cite{bonnor72} the name
electrically counterpoised dust, or ECD for short, is coined for the
first time for ${\sqrt G}\rho_{\rm m}=\epsilon \rho_{\rm e}$ dust,
i.e., for what we call and will always call, less contrively perhaps,
extremal charged dust, with the same acronym. Spherically symmetric
exact solutions are studied with the virtue that even when the
solutions are about to form a horizon the energy density $\rho_{\rm
m}$ is finite. Prolate solutions are also studied showing that in the
disk solution limit the energy density $\rho_{\rm m}$ is infinite,
naturally. Also, in particular, in \cite{bonnor75} several important
attributes of the solutions are perceived.  First, it is noticed that,
although no doubt, matter thus delicately balanced is rare, it is
physically possible and easily understood. Second, it is shown that
one can construct spheres of matter where infinite redshifts of light
emanating from the surface are attainable, whereas in an interior
Schwarzschild solution, say, only finite redshifts are
possible. Third, it displays exact solutions for spheroidal
configurations, and mentions that near the spheroidal horizon,
non-spherically symmetric features are filtered out. Fourth, it is
argued convincingly that these solutions are stable.  In
\cite{bonnor80} the study is interesting with queer results. First,
there is an incursion into solutions of the Newton-Coulomb with
an electrically charged fluid theory, where it is shown
that for a given Newtonian potential, call it $V$, there are equilibrium
non
Bonnor star solutions, not obeying the Majumdar-Papapetrou matter
condition, although these are singular. It also shows the analogue of
De and Raychaudhuri's theorem \cite{de-raichaudhuri}, i.e., that
Newtonian systems which do not obey the Majumdar-Papapetrou condition,
of the equality of mass and charge densities, and which have
equipotential surfaces, are singular. Second, in turning into general
relativity, with axial symmetry, spacetimes obeying the
Majumdar-Papapetrou condition are found, one of them being of physical
interest with positive energy density $\rho_{\rm m}$, and the others
of less interest.  In \cite{bonnor83,bonnor84} it is understood,
perhaps for the first time, that when the radius of the configuration
$r_0$ approaches the horizon radius, i.e., $r_0=M$, where $r$ is the
Schwarzschildean radial coordinate and $M$ the mass of the
configuration, the spacetime is somehow singular, being thus an idea
precursor of the concept of what a quasi-black hole is. In these works
the hoop conjecture is discussed and some lower bounds in connection
to it are given.  In \cite{bonnor98}, spheroidal bodies made of
extremal charged dust are studied in connection still with the hoop
conjecture and also with the isoperimetric conjecture, which says that
under certain general conditions $M\geq(A/16\pi)^{1/2}$, where $A$ is
the area of the apparent horizon, and $M$ the mass of the
configuration. In \cite{bonnor99} it is reinforced that spherically
symmetric configurations made of physically reasonable matter, though
admittedly not widely available, i.e., made of extremal charged dust,
yield solutions that can come arbitrarily close to the horizon of an
extremal black hole, and a general class of such solutions is
constructed by correct guessing.

Bonnor stars were studied further by other authors.  In
\cite{kleberlemoszanchin} a thick shell solution of Bonnor type was
found. In \cite{lemoszanchin1} it was noted that Bonnor star solutions
and gravitational magnetic monopole solutions have striking similar
properties, and a comparison of both solutions was performed and
discussed thoroughly. Previously, Lemos and Weinberg
\cite{lemosweinberg}, seeing in Bonnor's wake that these stars, which
are made of normal matter obeying the several important energy
conditions, can probe deeply the spacetime structure, proposed new
solutions, extended Bonnor star systems with a more sophisticated
density distribution asymptotic to an extreme Reissner-Nordstr\"om
solution, not needing any junction. Similar properties were found
for Bonnor stars properly said as well as for extended Bonnor
stars. Most notably, is the fact that at the threshold of the
formation of an event horizon the system displays a very peculiar
trait, instead of an extremal black hole one has a quasi-black hole,
with the formation of a quasihorizon instead of the usual event
horizon.  Although to external observers the system looks like an
extremal black hole, its internal properties are very different from
what one could expect in the case of a standard black hole. These
properties, along similar ones of gravitational magnetic monopoles and
glued vacuum solutions with shells,  have been analyzed in
\cite{lemoszaslavskii1,lemoszaslavskii2}. In \cite{horvat} other
attributes of these systems were explored.

\subsubsection{Further connections}

One can associate these Bonnor stars to several related topics.
(i) Both, astrophysically and physically, Bonnor stars are of
interest. On one hand, they can be realized if a gravitating sphere,
of neutral hydrogen which has lost a fraction 10$^{-18}$ of its
electrons, forms. On the other hand, they are supersymmetric solutions
of $N=2$ supergravity \cite{tod}, so are of interest in an elementary
particle context.
(ii) A matter always of maximal interest is the stability of the systems
one is considering, in this case, the Bonnor stars.  Interestingly
enough it was found, through different methods, that these stars are
neutrally stable.  Firstly, Omote and Sato \cite{omote} found this
stability criterion using both an energy method and a small adiabatic
radial oscillation method, results which were later confirmed in
\cite{glazer,anninos}.
(iii) When discussing static equilibrium configurations it is always
important to discuss the Buchdahl limits, where for instance for a
perfect fluid sphere, the star cannot reach beyond $r_0<9/8\,r_{\rm
Schw}$, where $r_0$ is the star radius, $r_{\rm Schw}$ is the
Schwarzschild radius, $r_{\rm Schw}=2GM$ \cite{buchdahl},
and $r$ is the Schwarzschildean radial coordinate.  Contrarily,
for Bonnor stars, stars made of extremal charged matter, the limits
are precisely the horizon radius as was first noted by Bonnor
\cite{bonnor53}-\cite{bonnor99}, and then in subsequent works
\cite{kleberlemoszanchin}-\cite{lemoszaslavskii2}, see also
\cite{yunqiang,mak,boehmer,giuliani,andreasson} for interesting
discussions on the Buchdahl limits for charged stars.
(iv) The hoop conjecture is relevant for these stars as was first
noticed by Bonnor \cite{bonnor83,bonnor84} (see also \cite{bonnor98}).
The conjecture states that a black hole forms when matter of mass $M$
is compacted within a given definite hoop, in \cite{thorne} taken to
be $\sim 4\pi GM$.  Later, it was shown that the hoop should be
reduced for extremal charged matter to $\sim 2\pi GM$
\cite{bonnor83,bonnor84}. However, it seems that systems like Bonnor
stars violate it, since no black hole forms ever, only a quasi-black
hole \cite{lemoszaslavskii1,lemoszaslavskii2}.
(v) A pertinent question, specially related to stars, is whether they can
form from gravitational collapse or not.  The issue of the collapse of
extremal charged dust solutions has not been studied in detail, see,
however, the interesting work of De \cite{ukde}.
(vi) Concerning the generalization of Bonnor stars to include pressure
terms, and thus go beyond dust matter there are some stimulating
developments. For instance, still within the Majumdar-Papapetrou
ansatz for the potential, $W=a_0-\epsilon\sqrt{G\,}\varphi$,
systems with pressure were studied by Ida \cite{ida00}, where one finds,
with an additional ansatz for the pressure, a Helmholtz type equation
which can be solved, in the case the pressure is zero see also
\cite{varela03}. These are thus extensions of Bonnor stars, stars that
include charged matter, non-extremal, and pressure. Extensions to
solutions with potentials different from the Majumdar-Papapetrou
potential, and even different to Weyl's potential, were done in
\cite{gautreau,guilfoyle}. These solutions include pressure and have
interesting structure.  Charged stars with pressure, were studied
numerically
in \cite{defeliceetal}, a paper that has attracted some attention,
where the limiting configuration is found to have mass equal to charge,
in appropriate units,
being thus a Bonnor star.  In \cite{malheiro} a set of static charged
solutions with pressure were studied and it was proposed that their
gravitational collapse would lead to the formation of a charged
Reissner-Nordstr\"om black hole.
(vii) There are many other
solutions of charged matter in various situations which have been
discovered throughout the years. Many of them are interesting and
would deserve a review, but there are too many to be quoted here, see
\cite{ivanov02} for a very partial list.
(viii) Charged gravitating solutions have been also used to study
Abraham type models for the electron, with and without Poincar\'e
stresses see, e.g, \cite{adm,katz,tiwari}, and also \cite{bonnor60}
and \cite{lemoszaslavskii2}.
(ix) A related issue to Bonnor star solutions and quasi-black holes is
the set of black holes devised by Bardeen \cite{bardeen}, in which the
interior to the horizon is nonsingular.  These solutions are
magnetically charged, instead of electrically charged, and have been
further explored in \cite{beato}.  The connection with the quasi-black
holes is that there is a theorem by Borde \cite{borde} which says that
if there is no singularity inside the event horizon then the regular
solutions have different inside and outside topologies.  Now, it is
not possible to put extremal charged dust, with positive rest mass,
inside an extremal black hole, \`a la Bardeen, a result first found
in $d$-dimensional studies (see below). So physical
(positive rest mass) Bonnor stars do not provide Bardeen like
solutions. On the other hand quasi-black holes are not true black
holes, but have a weird topology and properties
\cite{lemoszaslavskii1,lemoszaslavskii2}, approaching considerably the
topology change of Borde. For further connections of Bonnor stars and
quasi-black holes with other issues, such as no hair theorems, naked
black holes, objects that mimic black holes, and the entropy issue,
see \cite{lemoszaslavskii1,lemoszaslavskii2}.

\subsection{Higher dimensional analyses}

\subsubsection{Context}

The possibility of the existence of extra dimensions arise in several
theoretical schemes. In what is called a Kaluza-Klein unification
model, the unification idea has emerged first as a way of unifying the
gravitational and electromagnetic fields in five spacetime dimensions,
and later the gravitational and Yang-Mills fields in seven spacetime
dimensions. Within this idea the gravitational field in higher
dimensions gives rise to the gravitational field itself and the other
possible fields in four dimensions. Later, the Kaluza-Klein process
was enforced in theories which start from the outset in higher
dimensions, such as supergravity or string-M theory, which can have up
to eleven dimensions. In the course of reducing the dimensions to
four, a profusion of new fields materialize, see
\cite{bookwithoriginalpapers} for the original papers. These schemes
require that the extra dimensions are compactified in Planck size
manifolds, and so due to the lack of a properly accepted theory at
these scales it is very hard to do physics on the extra dimensions. It
has now appeared an idea that makes the higher dimensions large, when
compared to the Planck scale, which means, if correct, it can have
measurable consequences on current or near future experiments. By
postulating that the gravitational field propagates also in at least
extra three space dimensions, while electromagnetism and the standard
model fields propagate only in our universe, the brane, it is possible
to reduce the Planck scale to the electroweak scale and make the extra
dimensions large, of the order of hundredths of a centimeter or a bit
less. The hierarchy problem, of understanding the huge differences in
the gravity and electroweak scales, is now pushed into the acceptance
of the large extra dimensions (see \cite{antoniadis,arkmed,aaddvali}, 
see also \cite{randallsundrum} for possible developments).

Now, since within this arrangement gravity is an electroweak scale
phenomenon, so are black holes. Thus, for instance, by splashing
electrically charged particles together black holes or other
gravitational objects can be created in higher dimensions with the
charge remaining in the brane.  In scenarios with extra dimensions it
is thus important to study charged solutions in connection to the
formation of these tiny black holes because the charge and the
solutions suggest that the charge may halt gravitational collapse.
Solutions for charged objects in such a frame are certainly not
spherically symmetric, so not Reissner-Nordstr\"om, and at the moment
have not been found.  Nonetheless, it is certainly of interest to
consider spherically symmetric electrically charged solutions in
higher dimensions because, first such a study can give an idea of how
the existence of the charge influences the solution, and second,
other charged fields, analogous in many respects to the
electromagnetic field, may propagate in the higher dimensions, making
Maxwell electrically charged solutions prototype solutions.

In addition, related to studies on the role played by the
dimensionality of space on the laws of physics and its connection to
the anthropic principle, it has been shown that there are no Newtonian
solutions for stars supported by degenerate pressure in higher
dimensions, i.e., a higher dimensional self-gravitating Fermi gas
either collapses into a black hole or evaporates. Indeed, interesting
papers discussing degenerate stars, like white dwarfs and neutron
stars, in higher dimensions have appeared
\cite{bechhoeferchabrier,chavanis}.  In \cite{bechhoeferchabrier} a
complete heuristic study, following the original work of Landau (see,
e.g, \cite{landaulifshitz}), was performed. Then in \cite{chavanis},
the full study, following the original works of Chandrasekhar (see,
e.g, \cite{chandrasekhar}), was completed. The main conclusion is that
there are no Newtonian solutions for degenerate stars in higher
dimensions, thus no general relativistic solutions also, the Fermi
pressure energy cannot balance the gravitational energy.  Of course
this may not follow for other stars. Stars supported by classical gas
pressure may perhaps exist in higher dimensions, no conclusive study
has been presented so far.  Thus, it is of interest to show whether
stars, supported by electric repulsion, such as Bonnor stars, do have
solutions within Newtonian gravity and general relativity. In case
there are solutions, one shows by example, that the existence of stars
in higher dimensions depends both on the number of dimensions itself,
and on the underlying field content of the stars themselves.

It is thus important, for the reasons just raised,
to study Bonnor stars in higher dimensions, prior to
compactification of any sort.

\subsubsection{Vacuum Majumdar-Papapetrou solutions}

Electrically charged solutions in vacuum in $d$ dimensions within
Newton-Coulomb theory are a direct generalization from four
dimensions. Within Einstein-Maxwell theory the $d$-dimensional
solutions were found by Tangherlini \cite{tangherlini}, with a
prescient discussion on the physical laws and their
relationship to the three dimensionality of space.  These
solutions generalize the four dimensional Reissner-Nordstr\"om
solutions, and they also have the mass $M$ and the charge $Q$, as the
higher dimensional parameters, such that for $\sqrt{G_d\,}M<\epsilon
Q$ one has a charged naked singularity, for $\sqrt{G_d\,}M>\epsilon Q$
one has a Reissner-Nordstr\"om black hole, and $\sqrt{G_d\,}M=\epsilon
Q$ one has an extremal black hole, where $\epsilon=\pm 1$ depending on
the sign of the charge. Here $G_d$ is the $d$-dimensional Newton's
gravitational
constant, where in four spacetime dimensions we put $G_4\equiv G$ (see
Appendix \ref{appendixgd} for more on this).  If one takes Weyl's and
Majumdar's route into higher dimensional Einstein-Maxwell
theory, see now \cite{lemoszanchin2}, and seeks the initial ansatz
that the metric potential $W$, or its inverse $U=W^{-1}$, depends on
the electric potential $\varphi$, i.e., $W(\varphi)$ with
$g_{00}\equiv W^{2}$, one finds the relation $W^2 = \left(
a_0-\epsilon\sqrt{G_d}\,\varphi \right)^2 + b_0$, also independent of
the symmetry.  In the perfect square Majumdar-Papapetrou case, i.e.
$W=a_0-\epsilon\sqrt{G_d}\,\varphi$, one can also
show that specializing to spherical symmetry, the mass $M$ of the
solution is equal in appropriate units to its charge $Q$,
$\sqrt{G_d\,}M=\epsilon Q$. This makes contact with the Tangherlini
black holes through the extremal solution $\sqrt{G_d\,}M=\epsilon Q$,
although not through the other ones, since the Tangherlini family
generically does not admit a functional relationship between the
metric and the electric potentials. The complete understanding of a
single extremal Reissner-Nordstr\"om solution can also be achieved
through Carter-Penrose diagrams, and the complete understanding of the
vacuum Majumdar-Papapetrou solutions, with many extremal black holes
scattered around in $d$ dimensions was performed in \cite{myers}.

Moreover, it is interesting to note that if instead of working in
Einstein-Maxwell theory, one works in string-M theory or in
supergravity theory in eleven dimensions, there are
Majumdar-Papapetrou type solutions, in the sense that the attraction
due to the gravitational field is counter-balanced by the repulsion of
the charged field of the theory, see,  e.g, \cite{marolf},
as well as \cite{peet}, for reviews on this topic
(see also \cite{emparreal} for a review on black
hole and other solutions of higher-dimensional vacuum general
relativity and higher-dimensional supergravity theories).  In eleven
dimensions in string-M theory, there are two bosonic fields, the
metric and the $A_3$ form field which is a variant of the
electromagnetic field with a corresponding charge, and one fermionic
field. Thus the bosonic part is as simple as Einstein-Maxwell. One
finds that for the solutions to be purely bosonic one has to have that
the mass of the solution has to be equal to the $A_3$ charge,  in
appropriate units.  Note that this is the analogue of the extremality
bound for Reissner-Nordstr\"om black holes.  Solutions with mass equal
to charge are called BPS (Bogomolnyi-Prasad-Sommerfield)
spacetimes. The solutions are not point-like generically.  They are
brane like, and are called p-branes, or black p-branes, where a
zero-brane is a zero dimensional object like a black hole, a one-brane
is a string like a black string, a two-brane is a membrane and so on.
Indeed, in string-M theory, where supergravity in eleven dimensions is
a low energy theory, there are the M2-brane (a membrane,  i.e., a
two-brane electrically charged under $A_3$), the M5-brane (a
five-brane magnetically charged under $A_3$), the wave solution or
Aichelburg-Sexl metric, and the Kaluza-Klein monopole. All of these
are BPS, the last two having momentum which is a form of charge.
These solutions are best found and studied in isotropic, also
called harmonic, coordinates,
as is the case of Majumdar-Papapetrou solutions in general
relativity. One can then have, for instance, many M2-branes scattered
around, as one can have many black holes in the Majumdar-Papapetrou
case, since the charged field force still balances the gravitational
force. One can, in addition, combine the solutions with different
charge type, for instance a M2-brane with a M5-brane, with no analogue
in Majumdar-Papapetrou since here there is only one charge. Through
careful dimensional reduction, Kaluza-Klein or otherwise, these
solutions are also solutions of the reduced theories.  Usually the
branes in eleven dimensions are non singular and considered as
solitonic objects. But when one reduces to ten dimensions,
singularities in the solutions appear, in which case it is better to
consider the branes as coupled to extremal dust, in the place of the
singularities (see, e.g., \cite{marolf,peet}),  making thus the
consideration of extremal dust solutions in higher dimensions a
subject of interest.
It is also worth commenting that in string-M theory in eleven
dimensions one can also perform some brane engineering, by adding
together solutions of the same type of charge. It is common practice
to stack an array of M2 electrically charged branes, for instance, and
then take the continuum limit, or smear, the array in the correct
direction, yielding a new brane with a new dimension, see, e.g.,
\cite{marolf,peet}.  Of course one can also do the same type of
manipulation in Majumdar-Papapetrou general relativity. Draw an array
of equally sparse extremal black holes on a line, smear them together
correctly, and obtain a one dimensional extremal black string obeying
the $d$-dimensional Majumdar-Papapetrou equations.

\subsubsection{Beyond vacuum: dust Majumdar-Papapetrou solutions}

In $d$ dimensions, as in four, things become more interesting if one
goes beyond vacuum and puts an electrically charged fluid or matter
into the  Newtonian gravity or general relativistic systems of
equations.  Leaning on the general relativistic analysis of
Majumdar \cite{majumdar}, Lemos and Zanchin \cite{lemoszanchin2}
showed, for the special relation, or ansatz in this context, on the
metric inspired from the vacuum case, i.e.,
$W=a_0-\epsilon\sqrt{G_d\,}\varphi$, that the whole system
reduces to a single equation, a Poisson type equation, in which the
mass density $\rho_{\rm m}$ is equal to the charge density $\rho_{\rm
e}$ in appropriate units, $\sqrt{G_d\,}\rho_{\rm m}=\epsilon\rho_e$.
Thus, a basic feature of such a system is that, although being a
system containing charged matter, it is described by the metric
function $W$, the redshift function.  The electric potential $\varphi$
can then be found implicitly through the Majumdar-Papapetrou relation.
It is also possible to generalize to $d$ dimensions the theorem,
proved in four dimensions in general relativity in
\cite{de-raichaudhuri}, that, provided the pressure is zero and there
are no singularities in the distribution, the Majumdar-Papapetrou
ansatz $W=a_0-\epsilon\sqrt{G_d\,}\varphi$ and condition
$\sqrt{G_d\,}\rho_{\rm m}=\epsilon\rho_{\rm e}$ follow
\cite{lemoszanchin4deraichaudhuri}. One can then show that the
$d$-dimensional Newtonian limit follows, with the four dimensional
situation studied in \cite{bonnor80} being a particular case. Also
theorems with nonzero pressure \cite{guilfoyle} can be render into $d$
dimensions \cite{lemoszanchin4deraichaudhuri}, namely, that for
perfect fluid solutions satisfying the Majumdar-Papapetrou condition
the pressure is related to redshift function, as in the four
dimensional case \cite{guilfoyle}.

\subsubsection{Bonnor stars: junction of dust with vacuum
Majumdar-Papapetrou solutions}

In \cite{lemoszanchin2} it was proved that if the pressure is
functionally related to the redshift function, which in turn obeys the
Majumdar-Papapetrou relation for the potentials, then to have a
surface with zero pressure, i.e., a star, one has to have the pressure
equal to zero everywhere. This in turn means the star is a Bonnor
star, with a $d$-dimensional Majumdar-Papapetrou interior and a
$d$-dimensional extremal Reissner-Nordstr\"om exterior. This
result is valid within both Newtonian gravity and general relativity.
To discuss Bonnor star solutions in the spherically symmetric case is
the aim of this paper.  We show that spherically symmetric Bonnor
stars in $d$ dimensions have a number of interesting properties. 
In Newtonian theory their mass and radius may be arbitrary and the
object with the highest compression is a point electric mass, i.e., a
Newtonian singularity. In general relativity the stars can yield very
large redshifts and their exteriors can be made arbitrarily near to
the exterior of an extremal charged black hole. Even in these extremal
situations, many of their characteristics remain finite and
non-trivial. These extremal kind of $d$-dimensional systems are the
quasi-black holes, possessing quasihorizons, already mentioned.

\subsubsection{Further connections}

As in four dimensions, in $d$ dimensions one can try to
associate Bonnor stars to several related topics.
(i) Bonnor stars, or something related, in higher dimensions are of
interest in situations prior to compactification.  Since
astrophysically, the world is already compactified to four spacetime
dimensions, the main interest in these solutions is for high energy
physics, for instance in a large extra dimension scenario, where
charged configurations in higher dimensions can be of interest. It
would be of interest to know whether $d$-dimensional Bonnor stars, for
generic $d$, are supersymmetric solutions when embedded in some
supergravity theory.
(ii) Of course, the study of the stability of these higher dimensional
stars is important, although we do not do it here.
(iii) Buchdahl limits in higher dimensions have not been found
neither for uncharged nor for charged stars.
We are preparing such a study.
(iv) As far as we know, there is no discussion of
the hoop conjecture for objects in $d$ dimensions.
(v) In $d$ dimensions it is also important to understand if the
configurations under study can form from gravitational collapse.
Collapsing and static charged shells in $d$ dimensions within
Einstein-Maxwell theory  with an electrically charged fluid
have been analyzed in \cite{gaolemos}.  Static
shells, with vanishing pressure, in this context are
Majumdar-Papapetrou solutions. Collapsing charged shells in Lovelock
theory have been studied in \cite{gdiaslemos}.
(vi) One can also use the Majumdar-Papapetrou relation for the
potential $W=a_0-\epsilon\sqrt{G_d\,}\varphi$, and study
systems with pressure in much the same way as Ida \cite{ida00}. We
will not discuss this type of solutions in $d$ dimensions.
(vii) There are a few other, non Majumdar-Papapetrou type, solutions
of charged matter, see, e.g, the interesting ones discussed in
\cite{wolf,harko1}, where charged spheres with
specific distributions of matter, charge, and pressure were found.
(viii) Electron models in the molds of Abraham and Poincar\'e have not
been studied in $d$ dimensions.
(ix) It would be interesting to study Bardeen models and Borde's
theorem in $d$ dimensions.  An interesting result first derived in
\cite{gaolemos} is that for a shell in $d$ dimensions with positive
proper mass there is no static solution inside the event horizon, 
the result being valid if four dimensions as mentioned above, as well
as in $d>4$. This in some sense connects with Borde's theorem
\cite{borde}.

\subsection{Lay out}

We start by analyzing, in Section \ref{ndnewtoniangeneral0}, the
Newtonian theory for charged fluids in higher dimensions, looking for
static solutions. We verify in subsection \ref{ndnewtoniangeneral}
that if the condition $\sqrt{G_d\,}\rho_{\rm m}=\epsilon\rho_{\rm e}$,
with $\epsilon=\pm1$, is to be satisfied, then there are equilibrium
Bonnor star solutions in $d=(n+1)$-dimensional spherically
symmetric spacetimes, where $n$ is the dimension of the
space, see subsection \ref{ndnewtoniansolutions}. Bonnor stars of
Majumdar-Papapetrou general relativity are studied in Section
\ref{ddrelativisticgeneral0}. In subsection
\ref{ddrelativisticgeneral} we write the basic equations and
particularize them for spherical symmetry. Part of the subsection is
devoted to review the main properties of $d$-dimensional
spherically symmetric solutions.  A solution
of a $d$-dimensional Bonnor star is also analyzed in subsection
\ref{bonnorrel} in some detail.  Its generic properties are shown, as
well as its quasi-black hole  and  its quasi Newtonian limits. In
Sec. \ref{conclusions} we present final comments and conclusions.

\section{Newton-Coulomb theory with an electrically charged fluid
in $d$-dimensional spacetimes ($d=n+1$,
$n$ being the number of space dimensions):
Weyl and Majumdar-Papapetrou analysis, and Bonnor star solutions}
\label{ndnewtoniangeneral0}

In $d$-dimensional Newtonian gravity coupled to both Coulomb
electricity and a charged fluid matter, one can find solutions
representing
charged stars, where here $d$ is the number of spacetime dimensions,
with $d=n+1$, $n$ being the number of space dimensions. The dynamics
of such a kind of system is governed by the Euler equation, where the
gravitational and electric force fields are determined conjointly by
Newtonian gravity and Coulomb electricity.  The fluid can be in static
equilibrium even with zero pressure and stresses, because the electric
repulsion counterbalances the gravitational pull if the charge
density of the fluid, $\rho_{\rm e}$, equals its mass density
$\rho_{\rm m}$ in appropriate units, i.e., $\sqrt{G_d\,}\rho_{\rm
m}=\epsilon \rho_{\rm e}$, where $G_d$ is Newton's gravitational
constant in $d$
dimensions (see appendix \ref{appendixgd}),
and $\epsilon=\pm1$. This condition makes possible
to build a distribution of charged dust with any shape in neutral
equilibrium. Charged fluids with $\sqrt{G_d\,}\rho_{\rm
m}=\epsilon\rho_{\rm e}$ are called extremal charge dust fluids.  By
introducing a convenient boundary one turns the solutions into
stars. In this section we study some properties of these objects.  One
can also put some form of pressure, either positive or negative, into
these systems and find solutions which of course do not obey the
extremal condition. Solutions with pressure will not be considered
here.

\subsection{The gravitating Newtonian charged dust fluid, and
Weyl and Majumdar-Papapetrou type analysis}
\label{ndnewtoniangeneral}

\subsubsection{The gravitating Newtonian charged dust fluid}
\label{The charged fluid model}

We consider first the dynamics of a gravitating Newtonian charged
fluid in a $n=(d-1)$-dimensional Euclidean space according to the Euler
description. A dust fluid is completely specified by its velocity
vector, with components $v_i$, with $i=1,...,d-1$ (Latin indices run
through $1$ to $n=d-1$), and its matter density $\rho_{\rm m}$,
all being functions of the position vector represented
by spatial coordinates $r_i$, and of the universal time $t$.
Thus, $v_i = v_i(r_j,t)$ and $\rho_{\rm m}=\rho_{\rm m}(r_j,t)$.
The basic equations
governing the flow of a Newtonian fluid are the continuity equation,
which expresses mass conservation, and the Euler equation, which
expresses momentum conservation.  Consider a fluid element with mass
$dm = \rho_{\rm m} \, d{\cal V}$, in the $(d-1)$-dimensional space,
$d{\cal V}$ being the $(d-1)$-dimensional  space volume element.  Then, the
continuity and the Euler equations may be written as
\beqa
&& \frac{\partial \rho_{\rm m}}{\partial t} +
\nabla_i\left(\rho_{\rm m}\,
v^i\right)=0 \, , \label{continuity}\\
&&\rho_{\rm m} \frac{d  v_i}{ d t} =
F_i\, ,\label{euler}
\eeqa
respectively, where $d/dt\equiv \partial/\partial t + v^i\,\nabla_i$
is the convective temporal derivative, $\nabla_i$ is the
$(d-1)$-dimensional gradient operator, $F_i$ is the volumetric external
force acting upon the fluid element, and the sum convention on indices
is adopted.  The Newtonian systems we are interested in here are
gravitating charged fluids distributions in static equilibrium. The
fluid is then allowed to have some net electric charge, so that the
charge of a fluid element is $dq=\rho_{\rm e}\,d{\cal V}$, $\rho_{\rm
e}$ being the electric charge density of the fluid. Thus,
there are two independent forces acting on a fluid element,
the gravitational  and electrostatic forces.  Both these forces may be
derived from scalar potentials, $V$ and $\phi$, respectively, such
that one has
\beq
F_i=-\rho_{\rm m} \nabla_i\,V -\rho_{\rm e}\nabla_i\,\phi \, .
\label{forces}
\eeq
The gravitational potential $V$ is related to the mass density
$\rho_{\rm m}$ by
\beq
\nabla^2 V = S_{d-2}\,G_d\,\rho_{\rm m}\, , \label{gravitypoisson}
\eeq
while the electric potential $\phi$ is related to the charge density
$\rho_{\rm e}$ by
\beq
\nabla^2 \phi = - S_{d-2}\, \rho_{\rm e}\, , \label{electricpoisson}
\eeq
where the operator $\nabla^2$ is the Laplace operator in
$d-1$ space dimensions,
$S_{d-2}$ is the area of the unit sphere in $(d-1)$-dimensional space
given
by $S_{d-2}= 2\pi^{(d-1)/2}/\Gamma((d-1)/2)$, $\Gamma$ is the usual gamma
function, and $G_d$ is Newton's gravitational constant in
$d=n+1$ dimensions (see Appendix \ref{appendixgd} for the
definition of $G_d$).
$S_{d-2}$  reduces to $4\pi$ in four spacetime dimensions and Eqs.
(\ref{gravitypoisson}) and (\ref{electricpoisson})  are the natural
generalizations of the corresponding three-dimensional Poisson equations
for the potentials $V$ and $\phi$ to $(d-1)$-dimensional space.

We will consider only static systems, so all
quantities are functions of the $d-1$ space coordinates only, and the
fluid's velocity can be put equal to zero, $v_i=0$. Then the Euler
equation (\ref{euler}) for the charged fluid in static equilibrium reads
\beqa
\rho_{\rm m} \,
\nabla_i\, V
+ \rho_{\rm e}\,\nabla_i\,\phi
=0\, .
\label{eulerstatic}
\eeqa
Equations (\ref{gravitypoisson})-(\ref{eulerstatic}) are the
important equations for the problem.  Eqs.
\eqref{gravitypoisson}
and \eqref{electricpoisson} are the field equations that determine the
gravitational and the electric potentials once the mass and charge
densities are given, while Eq. (\ref{eulerstatic}) is the equilibrium
equation for the system.

\subsubsection{Weyl and Majumdar-Papapetrou type analysis}
\label{weylansatz_section}

\noindent

In vacuum, doing for Newtonian gravity what Weyl did for general
relativity \cite{weyl}, assume now an ansatz, i.e., a functional
relation, between the gravitational and the electric potential,
\beq
V=V(\phi)\, . \label{weylhyp}
\eeq
Eq. (\ref{weylhyp}) is the Weyl ansatz which implies that $V$ and $\phi$
have the same equipotential surfaces.  With this  ansatz, Weyl
originally worked out the Einstein-Maxwell vacuum equations that would
follow and found that the relativistic potential is a quadratic
function of the electric potential.  Doing the same here in Newtonian
gravity, we find that the ansatz \eqref{weylhyp},
in vacuum, $\rho_{\rm m}=0$ and $\rho_{\rm e}=0$,
when put into Eqs. \eqref{gravitypoisson} and \eqref{electricpoisson},
gives the following equation,
$
\left({V}^\prime\right)^2 \,\nabla^2\,\phi+ V'\,V''\,
\left(\nabla_i\,\phi\right)^2 =0\,,
$
where the prime stands for the derivative with respect to $\phi$.
Thus, since $\nabla^2\,\phi=0$ in vacuum, and
$\left(\nabla_i\,\phi\right)^2 \neq0$, it follows that $,V''=0$, i.e.,
$V(\phi) = a_0+ {\rm const}\times\phi\,$, where
$a_0$ is an arbitrary constant, that without loss of generality can be
put to zero. In addition, with our choice of units one has
that ${\rm const}=-\epsilon\sqrt{G_d}$. Thus,
\beq
V(\phi) = a_0 -\epsilon\sqrt{G_d}\,\phi
\label{weylsansatzpotentials}\, .
\eeq
This is the Weyl relation for the Newton-Coulomb theory in vacuum.
Following Majumdar \cite{majumdar} and Papapetrou \cite{papapetrou}
lead in general relativity, it is interesting to investigate the
consequences of the linear relation between electric and Newtonian
potentials, $V= a_0+{\rm const}\times\phi$, and see what happens in
presence of matter. Such a relation is a particular case of the Weyl's
general ansatz \eqref{weylhyp}, and is the same as in Eq.
\eqref{weylsansatzpotentials}, i.e., is the same as Weyl's relation
for vacuum. It is remarkable that the Majumdar-Papapetrou relation is
equivalent to Weyl's relation in Newton-Coulomb theory in vacuum,
while it is not so in general relativity. We use this relation
\eqref{weylsansatzpotentials} to treat also solutions with matter, as
has been done in general relativity \cite{lemoszanchin2}.

In matter, we work out the basic equations using the general form
\eqref{weylsansatzpotentials} and so generalize to higher dimensions
the analysis in four dimensions done by Bonnor \cite{bonnor80}.  We
show for $d-1$ space dimensions the interesting result that the
equality of mass and charge densities follows from the field equations
as long as there are no singularities within the charged matter
distribution (see Bonnor \cite{bonnor80} for Newtonian systems, and De
and Raychaudhuri \cite{de-raichaudhuri} for general relativistic
systems in four dimensions).  The basic equations
\eqref{gravitypoisson}, \eqref{electricpoisson}, and
\eqref{eulerstatic}, can be rewritten by taking the Weyl ansatz
\eqref{weylhyp} into account.  To begin with, it is convenient to
consider first Eq.  \eqref{eulerstatic}, which now reads
$\left(\rho_{\rm m}\,V^\prime +\rho_{\rm e}\right)\,\nabla_i\phi
=0\,.$ So, the two fields $V$ and $\phi$ have the same equipotential
surfaces.  Since we consider $\nabla_i\,\phi\neq0$,
Eq. \eqref{eulerstatic} is then equivalent to
$
\rho_{\rm m} \, V^\prime+ \rho_{\rm e}
=0\,,\label{eulerfunctional}
$
where again the prime stands for the derivative with respect to
$\phi$.  By substituting $\rho_{\rm m}$ from the previous equation
into Eq. \eqref{gravitypoisson}, one finds
$\left({V}^\prime\right)^2 \,\nabla^2\,\phi+ V'\,V''\,
\left(\nabla_i\,\phi\right)^2 = -S_{d-2}\, \rho_{\rm e}\, ,$ where we
made use of the assumption $V=V(\phi)$. Then, with the help of Eq.
\eqref{electricpoisson} one finds
$
\nabla_i\left(\sqrt{Z}\, {\nabla^i\phi}\right) = 0 \,,
\label{eulerfunctional3}
$
where $Z$ is defined as
$
Z \equiv {G_d - V^\prime}^2  \, .
\label{Zdef_newt}
$
Now, in order to have a nonsingular solution with closed boundary it
is required that $Z=0$, or equivalently, $(V')^2={G_d}$. All
equilibrium solutions with $(V')^2\neq G_d$ with a closed
equipotential hypersurface $S$ have a singularity within $S$. The most
physically interesting solutions are then those for which
$(V')^2={G_d}$.  Therefore, the Majumdar-Papapetrou relation for the
Newton-Coulomb theory with a charged dust fluid is
$
V(\phi) = a_0-\epsilon\sqrt{G_d}\,\phi\,,
\label{mpconditionwithp_newt}
$
where $ \epsilon \equiv \pm 1$, implying, after
Eq. \eqref{weylsansatzpotentials}, that the $\rm const$ 
appearing before the potential $\phi$ is indeed 
$-\epsilon\sqrt{G_d}$. Thus,
for the Weyl relation or Majumdar-Papapetrou relation (they are the
same here), Eq. \eqref{weylsansatzpotentials}, with the equations
$\rho_{\rm m} \, V^\prime+ \rho_{\rm e}=0$ and
$V(\phi) = a_0-\epsilon\sqrt{G_d}\,\phi$ derived above, gives
$
\rho_{\rm e}=\epsilon \sqrt{G_d\,}\,\rho_{\rm m}\,
\label{MPNewtonian}
$.
This last equation is the Majumdar-Papapetrou condition in Newtonian
gravity. Observe that the relation between the potentials we call
Majumdar-Papapetrou relation and the relation between the densities we
call Majumdar-Papapetrou condition. In the case
the Majumdar-Papapetrou condition holds, distributions of
charged dust of any shape can be put in equilibrium.  All the
quantities can now be given.  Once the mass density $\rho_{\rm m}$ is
given, the gravitational potential is determined by the Poisson
equation \eqref{poissonfinal} and all the other quantities,
including the electro-gravitational Newtonian spacetime
structure, and possible singularity structure, follow from
$V$ and $\rho_{\rm m}$.  The resulting system of equations can be put
in the form
\beqa
&& \nabla^2 \, V = S_{d-2}\,G_d\,  \rho_{\rm m} \, ,
\label{poissonfinal}\\
&& \phi = - \frac{\epsilon}{\sqrt{G_d}}\,\,V \,, \label{electricfinal}\\
&&\rho_{\rm e}=\epsilon\sqrt{G_d\,}\,\rho_{\rm m}\, .\label{chargefinal}
\eeqa
where the zero points of the potentials are suitably chosen.

\subsection{Spherical $d$ spacetime
($n$ space) dimensional Newtonian Bonnor star solutions}
\label{ndnewtoniansolutions}

\subsubsection{Equations in spherical coordinates}
We now assume the mass distribution is spherically symmetric, in which
case all the dynamical variables and fields depend only on the radial
coordinate $r$ in $(d-1)$-dimensional space.  Our interest here is in
spherical solutions to equations
\eqref{poissonfinal}-\eqref{chargefinal}.  First we define the mass
$m(r)$ and the electric charge $q(r)$ inside a sphere of radius $r$,
respectively, as
\beqa
m(r)=S_{d-2}\int_0^r{\rho_{\rm m}(r)\,  r^{d-2}dr}\,,
\label{masseq}
\eeqa
\beqa
q(r)= S_{d-2}\int_0^r{\rho_{\rm e}(r)\, r^{d-2}dr}\,.
\label{chargeinside}
\eeqa
Equations \eqref{poissonfinal}-\eqref{chargefinal} are then
conveniently written explicitly in terms of the radial coordinate as
\beqa
&&\frac{d V(r)} {dr} = G_d\, \frac{m(r)}{r^{d-2}}\, ,\label{gravityeq}\\
&&\frac{d \phi(r)}{dr}=-\frac{q(r)}{r^{d-2}}\, , \quad {\rm or}\quad
\phi(r) = - \frac{\epsilon}{\sqrt{G_d}}\,\,V(r) \,,
\label{phieqagain}\\
&&q(r)=\epsilon \sqrt{G_d\,} m(r)\,,
\label{mq}
\eeqa
where the zero points of the potentials were suitably chosen.
In Eq. \eqref{gravityeq} there is also a term ${C_0/ r^{d-2}}$
which we have put to zero, without loss of generality, i.e., $C_0=0$.
This term can be included in the term $G_d\,
{m(r)/ r^{d-2}} $ by an appropriate choice
of the function $m(r)$.

\subsubsection{Solutions}
\label{newtoniansolutions}

{\it\small (a) Electrovacuum solutions in $d=n+1$ spacetime dimensions}

The solution to Eq. \eqref{gravityeq} in vacuum is
\begin{eqnarray}
&&V=-\frac{1}{d-3}\,\frac{G_d\, M}{r^{d-3}} \,, \label{vofr}\\
&& M = {\rm constant}\, , \label{Mnewton}
\end{eqnarray}
with $M$ representing the Newtonian mass of the source.
To complete the solution one must give the electric potential $\phi$,
which
is obtained from Eq. \eqref{phieqagain},
$\phi = (d-3)^{-1}\,Q/r^{d-3}$, with $Q$ being the total charge
 of the source, which in turn satisfies Eq.\eqref{mq},
$Q = \epsilon\sqrt{G_d} M. $ These two equations, together with
 Eqs. \eqref{vofr} and \eqref{Mnewton}, form the set of
equations corresponding to the solution of a Newtonian
Majumdar-Papapetrou  vacuum system in $n=d-1$ space
dimensions. Such a set of solutions also follows from the Poisson
equations in which the mass and charge densities are Dirac delta
functions, $\rho_{\rm m}(r)= M\, \delta(r)$ and $\rho_{\rm e}(r)=
Q\,\delta(r)$, and $Q =\epsilon \sqrt{G_d\,}\, M$.

\vskip 0.1cm
{\it\small (b) Newtonian Bonnor star solutions in $d=n+1$ spacetime
dimensions ($n$ space dimensions)}
\label{ndnewtonianstars}

Now we find a class of solutions to the Newton-Coulomb system with
electrically charged fluid matter,
considering the Majumdar-Papapetrou relation
(\ref{weylsansatzpotentials}) (which in the Newtonian case is also
Weyl's relation) and a spherically symmetric distribution of
matter. Upon joining this class of solutions to an external vacuum we
obtain Bonnor stars in the Newtonian theory.  The relevant equations
are the ones presented in system \eqref{masseq}--\eqref{mq}.

Let us call $r_0$, the radius of the star.  Physical conditions
require the mass density to be a continuous function with a finite
value at the center of the star.  One can choose a mass density
function $\rho(r)$ and the remaining functions are then obtained by
integrating the appropriate equations.  For instance, one can give
$\rho_0\,\,({r}/{r_0})^\alpha$, for $0\leq r\leq r_0$, and make it
zero in all the exterior region for $r>r_0$. Integration of the
Poisson equation \eqref{gravityeq} gives a power law function for the
potential, $a_1\,r^{\alpha+2} + {a_2}/{r^{d-2}} +a_3$, where the
constant $a_2$ is made equal to zero in order to avoid a singularity
at $r=0$, and the constant $a_3$ is fixed by the matching conditions
at $r=r_0$.  Alternatively, instead of giving $\rho_{\rm m}(r)$, one
can choose a potential $V(r)$ satisfying reasonable boundary
conditions, and then obtaining the other functions from it.  This is
the simplest route, the one we follow here.  We can choose the
following interesting potential, for the interior $V_{\rm i}(r)$,
given by $ V_{\rm i}(r)= c_0 + c_1\, ({r}/{r_0})^\alpha +
c_2\,({r}/{r_0})^\beta$, for $r\leq r_0$, where $\alpha$ and $\beta$
are arbitrary constant parameters, possibly satisfying some
restrictions.  The other constants, $c_0$, $c_1$ and $c_2$, are fixed
by imposing appropriate matching conditions at the surface of the
star, $r=r_0$.  One can impose that the potentials are $C^1$ functions
at $r_0$, which is usually done in order to simplify the calculations.
This means continuity of the potential and continuity of the
gravitational field strength.  Then, in this case, the density has a
finite discontinuity at the boundary, falling from some finite value
just inside matter to zero just outside.  Also through the junction
conditions above, one can find the constants $c_0$, $c_1$, and $c_2$,
with one of them arbitrary, $c_1$ say. Here we want to go a step
further and impose that the potentials are $C^2$ functions at $r_0$,
i.e., continuity of the potential, continuity of its first derivative,
and continuity of its second derivative.  Continuity of the first
derivative of the potential means that the gravitational field
strength is continuous, and continuity of the second derivative means
that the mass density at the surface of the star is continuous with
zero value.  Continuity of the potential gives 
$
V_{\rm i}(r_0)= V_{\rm e}(r_0)= - (d-3)^{-1}
{G_d\,M}/{r_0^{d-3}}\, ,
$
where $V_{\rm e}(r)= -(d-3)^{-1} G_d\,{M}/{r^{d-3}}$
 is the Newtonian potential in the exterior region, $r\geq r_0$,  with
$V_{\rm e}$ zero at
infinity,
and $M$ is the total mass of the star.
Continuity of the gravitational field strength gives
$
V'_{\rm i}(r_0)= V'_{\rm e}(r_0)={G_d M}/{r^{d-2}} \, .
\label{matchfield}
$
Continuity of the mass density at the surface of the star gives
$
\rho_{\rm m}(r_0)=0 \, .
$
With these choices, the spherical Newtonian star
is described by the following potential
\beq
V= \left\{\begin{aligned}
  V_{\rm i} =&
-\frac{G_d}{d-3}\frac{ M}{r_0^{d-3}}\left(
1 + \frac{(d-3)(\beta+d-3)}{\alpha(\beta-\alpha)}
\left[1- \left(\frac{r}{r_0}\right)^{\alpha}\right] + \right.
 &  \\
 &  \left.-\frac{(d-3)(\alpha+d-3)}{\beta(\beta-\alpha)}\left[1-
\left(\frac{r}{r_0}\right)^{\beta}\right] \right) ,
& r\leq r_0\, ,  \label{sphericalpotential}    \\
 V_{\rm e}=& -\frac{G_d}{d-3}\frac{M}{ r^{d-3}} \, , & r>r_0 \, ,
\end{aligned}\right.
\eeq
and by the following mass density
\beqa
\!\!\!&\rho_{\rm m}& =\left\{\begin{aligned}
&  \frac{(\alpha+ d-3)(\beta+d-3)}
{(d-3) (\beta-\alpha)}\, \frac{M}{S_{d-2}\, {r_0}^{d-1}}
\left[\left(r\over r_0\right)^{\alpha-2} -
\left(r\over r_0\right)^{\beta-2}\right]  , & r\leq  r_0\, , \\
& 0\, , & r>r_0\, . \end{aligned}\right.
\label{sphericalmass}
\eeqa
In these equations $M$ is the mass and $r_0$ is the radius of the
star, with $M$ being obtained from Eq. \eqref{masseq} with $r=r_0$, i. e.,
$M= m(r_0)$, and $\alpha$ and $\beta$ are arbitrary constant parameters
satisfying the restrictions $\alpha\geq2$ and $\beta\geq 2$. In
addition, the parameters $\alpha$ and $\beta$ must be different from each
other, $\alpha\neq\beta$, in order that
the mass density be finite at $r=0$, and the other functions that follow
from it be also finite there.
Such conditions ensure also the positivity
of the mass density.
Note that the quantity $M/(S_{d-2}\, {r_0}^{d-1})$ appears naturally,
indeed in Newtonian theory one can define the mean density of the matter by 
${\bar\rho_{\rm m}}
=(d-1)\,M/\left(S_{d-2}\, {r_0}^{d-1}\right)$.
The other quantities, $\phi$ and $\rho_{\rm e}$,
are obtained by substituting the expression for the gravitational
potential
and for the mass density given in Eqs.
\eqref{sphericalpotential}-\eqref{sphericalmass}
into Eqs. \eqref{electricfinal} and
\eqref{chargefinal}, respectively.
\begin{figure}[h]\begin{center}
\includegraphics[scale=.8]{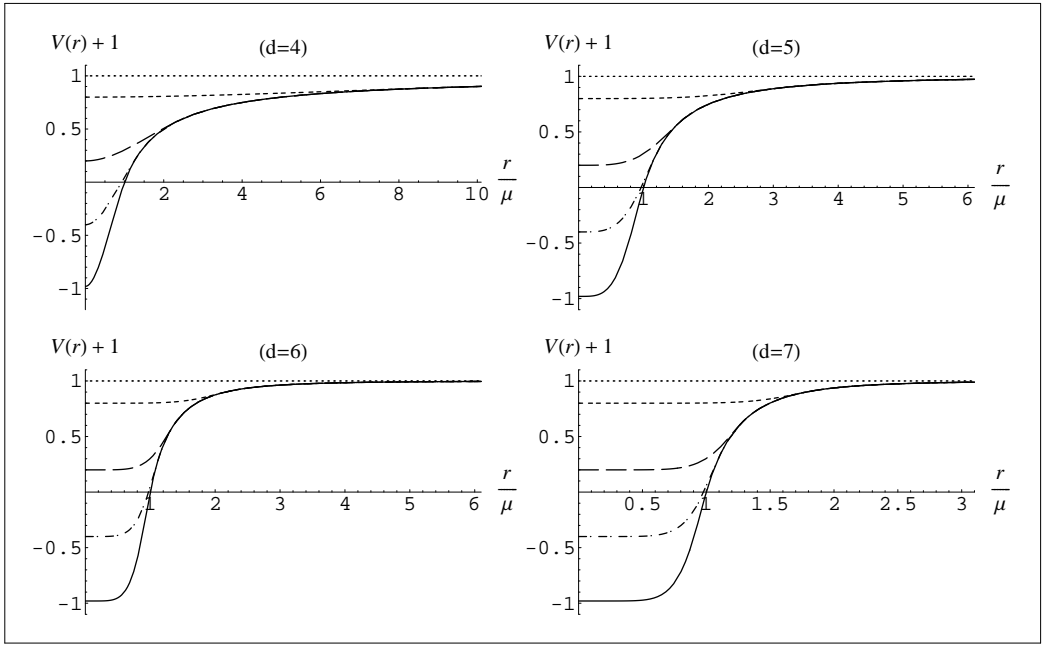}
\caption{The rescaled Newtonian potential $V(r)+1$ as a function of
$r/\mu$, where $\mu\equiv (G_dM/(d-3))^{1/(d-3)}$, for four spacetime
dimensions, $d=n+1$, $d=4$ (top-left panel), $d=5$ (top-right panel),
$d=6$ (bottom-left panel) and $d=7$ (bottom-right panel), and for four
different values of the parameter $a$.  The lowest, solid, curve is
for $a=1$, the dot-dashed line is for $a=0.7$, the dashed line is for
$a=0.4$, and the dotted line for $a=0.1$.}
\label{figure_VN} \end{center}
\end{figure}
So the class of Bonnor stars is defined essentially by equations
\eqref{sphericalpotential}-\eqref{sphericalmass}, through
the parameters $G_d$, $M$, $r_0$, $d$, $\alpha$ and $\beta$.
In the analysis of these Bonnor stars,
an important parameter appears,
the $d$-dimensional generalization of the mass to radius ratio of the
star,
\beqa
a= \frac{G_d}{d-3}\frac{M}{{r_0}^{d-3}}\, .
\label{parameter-a}
\eeqa
It measures how compact is the star, and is a free parameter in the
model.  Taking $M$ as a fixed parameter, different stars are
parameterized by different values of $a$, which means different values
of the radius $r_0$.  There are no constraints on $a$ for Newtonian
stars, it can vary from $0$, a highly dispersed star, to $\infty$, a
point mass, i.e., the limiting configuration here is a Newtonian
singularity at $r=0$ obeying the Majumdar-Papapetrou condition
$Q=M$. As we will see, in the relativistic case $a$ cannot be larger
than unity (see also
\cite{kleberlemoszanchin,lemoszanchin1,lemosweinberg}).

\begin{figure}[h]\begin{center}
\includegraphics[scale=0.8]{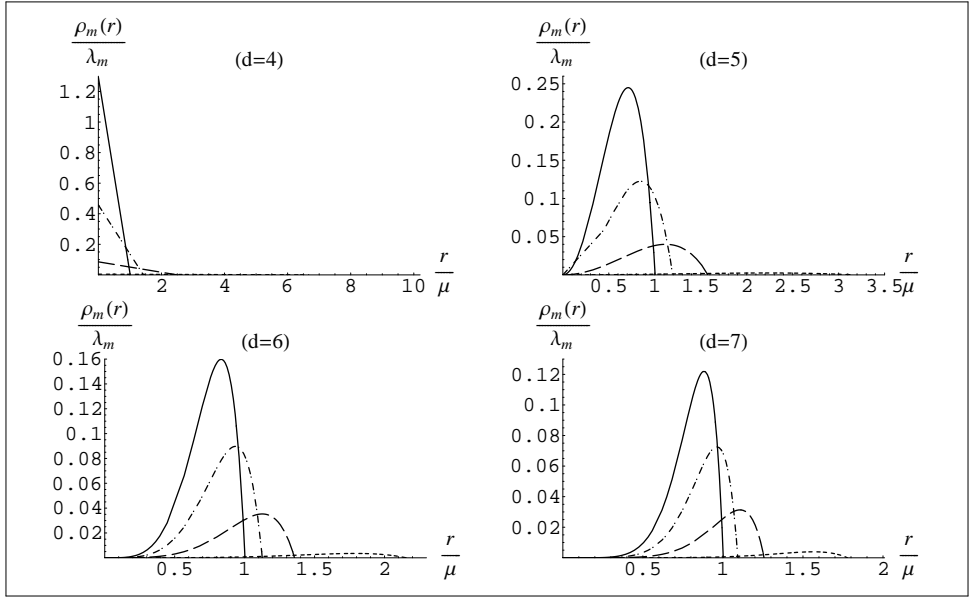}
\caption{The normalized Newtonian mass density $\rho_{\rm
m}(r)/\lambda_{\rm m}$ as a function of $r/\mu$, where $\lambda_{\rm
m}=\frac{12}{d-1}\bar\rho_{\rm m}$, $\bar\rho_{\rm m}$ being the
average density (see text), and $\mu\equiv (G_dM/(d-3))^{1/(d-3)}$,
for four different spacetime dimensions ($d=n+1=4$, $5$, $6$, $7$, as
indicated) and for four different values of the parameter $a$..  The
solid line is for $a=1$, the dot-dashed curve is for $a=0.7$, the
dashed line is for $a=0.4$, and the (lowest) dotted line is for
$a=0.1$.  The normalized Newtonian mass density $\rho_{\rm
m}(r)/\lambda_{\rm m}$ goes to zero at the surface of the star,
defining thus the radius $r_0$ in each plotted case.}
\label{figure_rhoN}
\end{center}
\end{figure}

The relevant functions $V(r)+1$, $\rho_{\rm m}(r)$, $\phi(r)$, 
and $\rho_{\rm e}(r)$,
given in terms of the coordinate $r$ follow from the
above relations. They are dependent on the variable $r$,
and also depend on two other arbitrary parameters, the mass and the
radius of the star, $M$ and $r_0$, respectively. Instead of writing
the explicit form of such functions, it is more
convenient to plot them for several choices of parameters. In the
calculations we normalized the coordinate $r$ to the mass
parameter $\mu=({G_d\,}M/({d-3}))^{1/(d-3)}$ which was kept fixed. In
fact, the important parameter to this end is the mass to radius ratio
$a$, given by Eq. (\ref{parameter-a}). 
The function $V(r)+1$: The behavior of the rescaled potential $V(r)+1$
as a function of the rescaled coordinate $r/\mu$, 
for four different values of $a$ ($a=0.1$,
$a=0.4$, $a=0.7$, and $a=1$), and in four different spacetime
dimensions ($d=4$, $5$, $6$, $7$) is shown in Fig. \ref{figure_VN}. We
plot the rescaled function $V(r)+1$ instead of simply $V(r)$ for
direct comparison with the relativistic case studied later.  Now, the
parameters $\alpha$ and $\beta$ in the solution
\eqref{sphericalpotential}-\eqref{sphericalmass} are free
parameters. We have chosen them so that $\beta=3\alpha/2=3(d-3)$,
which is a convenient choice when one studies the counterparts of
these solutions in general relativity. With this choice, the form of
the curves depends on the number of spacetime dimensions $d$ and on
the parameter $a$ alone.  Note that all the interior functions $V_{\rm
i}(r,a)$ match the exterior solutionn $V_{\rm e}(r)+1=
1-G_dM/((d-3)r^{d-3})$, each one at a different value of $r_0$. The
reason for that is because the change of $a$ is made by keeping the
mass of the star fixed, while $r_0$ varies accordingly.
The function $\rho_{\rm m}(r)$: 
Another quantity of interest is the mass density $\rho_{\rm m}(r)$.
In Fig. \ref{figure_rhoN} we plot $\rho_{\rm m}(r)/{\lambda_{\rm m}}$
as a function of the normalized radial coordinate $r/\mu$.  The
density $\lambda_{\rm m}$ is defined as $\lambda_{\rm m}=
\frac{(\alpha+d-3)(\beta+d-3)}{(d-1)(d-3)(\beta-\alpha)}\bar\rho_{\rm
m}$, where the mean density $\bar\rho_{\rm m}$ is given by
$\bar\rho_{\rm m}={(d-1)M}/(S_{d-2}\, r_0^{d-1})$. For our choice of
parameters, $\beta=3\alpha/2=3(d-3)$, 
one has $\lambda_{\rm m}=\frac{12}{d-1}\bar\rho_{\rm m}$.  
It is seen that $\rho_{\rm m}(r)$ is
finite at $r=0$. In fact,
with our choice, $\rho_{\rm m}$ vanishes at $r=0$ for all
$d>4$. In addition it goes to zero at the surface of the star, 
defining thus the radius $r_0$ in each plotted case. 
The behavior of the potential $\phi(r)$ is
simply given by $\phi(r)=-(\epsilon/\sqrt{G_d})\,V(r)$, and it is not
plotted.  The behavior of the charge density is $\rho_{\rm e}(r)=
\epsilon\,\sqrt{G_d}\,\rho_{\rm m}$, and it is not plotted.  Note that
the potentials, $V$ and $\phi$, are $C^2$ functions of $r$, so that
the corresponding field strengths are continuous ($C^1$ functions, in
fact) through the surface of the star. The mass and charge densities,
$\rho_{\rm m}$ and $\rho_{\rm e}$, are $C^0$ functions vanishing at
$r=r_0$.  When $r_0\rightarrow0$ one obtains a point charge with a
central Newtonian, mild, singularity.  It is mild because it is not a
nasty spacetime singularity, it is a matter singularity only.

\section{Einstein-Maxwell theory  with an electrically charged fluid
in $d$ spacetime
dimensions ($d\!=\!n\!+\!1$): Weyl and Majumdar-Papapetrou analysis,
and Bonnor star solutions}
\label{ddrelativisticgeneral0}

In $d$-dimensional general relativity coupled to both Maxwell
electromagnetism and a charged fluid matter one can also find
solutions representing charged stars.  The fluid can be in static
relativistic equilibrium if it is made of extremal matter, where the
electric repulsion from the charge density of the fluid, $\rho_{\rm
e}$, counterbalances the gravitational pull from its mass density,
$\rho_{\rm m}$, in appropriate units, i.e., $\sqrt{G_d\,}\rho_{\rm
m}=\epsilon\rho_{\rm e}$.  Thus, relativistic Bonnor stars in $d$
dimensions can also be constructed. Within general relativity the
behavior and properties of these solutions is much richer, allowing
the possibility of quasi-black hole behavior, for a sufficient compact
object, rather than the point like dull singularity of Newtonian
objects. In this section we study some properties of relativistic
charged fluids in the context of a Majumdar-Papapetrou analysis, and
the corresponding Bonnor stars.

\subsection{The gravitating relativistic charged dust fluid, and
Weyl and Majumdar-Papapetrou analysis}
\label{ddrelativisticgeneral}

\subsubsection{The relativistic gravitating charged dust fluid}
\label{relativisticmodel}

With the aim of finding exact solutions for $d$-dimensional
Bonnor stars we firstly write the basic equations
for the Majumdar-Papapetrou systems and analyze their
general properties in brief.
In the following sections we particularize for spherically symmetric
spacetimes, show a particular solution and study it in some
detail.

The general relativistic analog of the Newtonian charged fluid
discussed in the preceding section was considered in
\cite{lemoszanchin2}. Such a relativistic system is described by the
$d$-dimensional Einstein-Maxwell with an electrically charged fluid
system of equations which read (we use units such that $c=1$),
\begin{eqnarray}
& &G_\munu = \frac{d-2}{d-3}S_{d-2}
G_d\left(E_\munu+T_\munu\right)\, ,
\label{einsteqs}\\
& & \nabla_\nu F^\munu = S_{d-2} J^\mu\, , \label{maxeqs}
\end{eqnarray}
with $G_\munu$, being the Einstein tensor, such that
$G_\munu=R_\munu-{1\over2}g_\munu R$, $R_\munu$ being the Ricci tensor
and $R$ the Ricci scalar.  The right hand side of Eq. \eqref{einsteqs}
bears a universal constant $G_d$, which in four dimensions corresponds
to the Newton's gravitational constant (see Appendix \ref{appendixgd}
for the definition of $G_d$).  $S_{d-2} =
2\pi^{(d-1)/2}/\Gamma((d-1)/2)$, where $\Gamma$ is the usual gamma
function, and the whole factor $(d-2)\,G_d S_{d-2} /(d-3)$ corresponds
to the $8\pi G$ term in four dimensions.  The electromagnetic
energy-momentum tensor, $E_\munu$, is given by
\beq
E_\munu= \frac{1}{S_{d-2}}\left(
{F_\mu}^\rho {F_\nu}_\rho -\frac14\,g_\munu
F_{\rho\sigma}F^{\rho\sigma}\right),
\label{emunu}
\eeq
where $F_\munu \equiv \nabla_\mu A_\nu -\nabla_\nu A_\mu$,
$A_\mu$ is the electromagnetic gauge field, with $\nabla_\mu$ being
the covariant derivative.   $J_\mu$, in Eq. \eqref{maxeqs},
is the current density
\beq
J_\mu = \rho_{\rm e} u_\mu\,,
\label{current}
\eeq
where $\rho_{\rm e}$ is the charge density and
$u_\mu$ is the velocity of the fluid in the
$d$-dimensional spacetime with $g_\munu
u^\mu u^\nu =-1$.
Finally,
$T_\munu$ is the the matter energy-momentum tensor for dust given by
\beq
T_\munu=\rho_{\rm m}u_\mu u_\nu\,,
\label{fluidemt}
\eeq
with $\rho_{\rm m}$ being the energy density of the fluid.
In all the above definitions, Greek indexes
$\mu,\nu$, etc., run from $0$ to $d-1$, where 0 represents the time,
and the other $d-1$ coordinates are spacelike.

It is assumed the spacetime is static, in which case the metric can be
written in the form
\begin{equation}
ds^2 = - W^2 dt^2 + \frac{1}{W^\frac{2}{d-3}}h_{ij}dx^idx^j\, ,
\label{metric}
\end{equation}
where Latin indices run from $1$ to $d-1$, $h_{ij}$ is the
metric in $(d-1)$-dimensional space,
and  $W$ is a function of the spacelike coordinates $x^i$ only.
The four-velocity and the gauge field are then given respectively by
\beq
u_\mu =  W\, \delta_\mu ^0\, ,
\label{umu}
\eeq
and
\beq
A_\mu = - {\varphi}\,\delta_\mu ^0\, ,
\label{amu}
\eeq
where the electric potential $\varphi$ is also a function of the space
coordinates alone. (Note that in the definition of $A_\mu$
we have put a minus sign in front of $\varphi$. Although not the
usual choice, this is the useful choice to compare with
the Newtonian case.)

From the Einstein-Maxwell with a charged dust fluid
equations one obtains the
following equations for $W$ and $\varphi$
\beqa
& &\nabla^2 W -\frac{1}{W}\left(\nabla_i W\right)^2
= \frac{G_d}{W}\,\left(\nabla_i\varphi\right)^2+S_{d-2}\, G_d\,
\,{W^{{d-5}\over {\rm d}-3}}
\rho_{\rm m} \, ,
\label{gravitypoissonrel}\\
& &\nabla^2\varphi =2{\nabla_i W\over W} \nabla^i\varphi-
S_{d-2}W^{{d-5\over {\rm d}-3}} \,\rho_{\rm e}  \, ,
\label{electricpoissonrel}
\eeqa
where $\nabla_i$ stands for the covariant derivative with respect to the
space metric $h_{ij}$.
Making now the connection to the Newton-Coulomb theory  with
a charged dust fluid, one may say that Einstein-Maxwell
with a charged dust fluid equations, Eqs. \eqref{gravitypoissonrel}
and
\eqref{electricpoissonrel}, correspond to the Poisson equations for
the gravitational and electric potentials, Eqs. \eqref{gravitypoisson}
and \eqref{electricpoisson}, respectively. Moreover, continuity and
Euler equations \eqref{continuity} and \eqref{euler} are, in certain
sense, analogous to the relativistic conservation equations
$\nabla_\mu E^\munu+\nabla_\mu T^\munu = 0\, , \label{conservationeqs}
$ which in turn also follow from the general relativity equations.
In the present case one has
\beq
\rho_{\rm m}\,\nabla_i\, W+\rho_{\rm e}\,\nabla_i\,\varphi=0\, ,
\label{conservationeq1}
\eeq
for the conservation equation,
which has the same form as Eq. (\ref{eulerstatic}).

\subsubsection{Weyl and Majumdar-Papapetrou analysis}
\label{weylsansatzrelativistic}

In vacuum, the generalization of the
Majumdar-Papapetrou system to $d$-dimensio\-nal spacetimes was done
in \cite{lemoszanchin2} and, for completeness, we summarize the main
properties of such systems here. Following the lines of that work but
changing the strategy in order to compare the present analysis to the
Newton-Coulomb case of previous sections, we assume there is a
Weyl-type functional relation between the metric potential $W$ and the
relativistic electric potential $ \varphi$
\beq \label{weylsansatzrel}
W=W(\varphi) \, .
\eeq
This is the relativistic Weyl's ansatz.
For the sake of comparison to the Newtonian case, let us review here
the main consequences of the last equation.  For the
vacuum case, $\rho_{\rm m}=\rho_{\rm e}=0$. So,
Eqs. \eqref{gravitypoissonrel} and \eqref{electricpoissonrel} can be
combined to yield $\left(\nabla_i\varphi\right)^2 (W\,W^{\prime\prime}
+ {W^\prime}^2- WG_d)=0 \, .$ Since $(\nabla_i\varphi)^2 \neq 0$, this
equation implies in $WW^{\prime\prime} + {W^\prime}^2- WG_d=0,$ which
integrates to
$
W^2 = \left( a_0-\epsilon\sqrt{G_d}\,\varphi \right)^2 + b_0
\, ,
\label{weylsansatzpotentialsrel}
$
where $a_0$ and $b_0$ are integration constants. This form of the metric
potential $W$ is known as the Weyl potential or, in our context,
the Weyl relation. Moreover, in the particular
case where $b_0=0$, $W^2$ assumes the form of a perfect square so that
\beq
W =  a_0 -\,\epsilon\sqrt{G_d}\,\varphi
\, ,
\label{MPsansatzpotentialsrel}
\eeq
where $\epsilon = \pm 1$, and without loss of generality we kept
the plus sign when taking the square root of $W^2$. In general
relativity, this form of $W$ is known as the Majumdar-Papapetrou
potential, and one usually refers to Eq.
\eqref{MPsansatzpotentialsrel} as the Majumdar-Papapetrou relation.

In matter, we now render into $d$ dimensions
De and
Raychaudhuri's theorem \cite{de-raichaudhuri} (see
\cite{lemoszanchin4deraichaudhuri}
for the generalization of it for systems with pressure).
To begin with,  one substitutes of Eq. \eqref{weylsansatzrel}
into the conservation equation \eqref{conservationeq1} and finds
$
\left(\rho_{\rm m} W'+\rho_{\rm e} \right) \nabla_i\varphi=0 \,
$
which, for $\nabla_i \varphi\neq 0$,  is then equivalent to
\beq
\rho_{\rm m} W' + \rho_{\rm e} =0 \, ,
\label{conserveq2}
\eeq
where the prime denotes derivative with respect to $\varphi$. This is
the general relativistic analog to the equilibrium equation of
Newtonian theory, cf. the equation $\rho_{\rm m} \, V^\prime+
\rho_{\rm e} =0$ derived in Sec. \ref{weylansatz_section}.
Using Eq. \eqref{conserveq2}, it is also readily seen that, together with
Eqs. \eqref{gravitypoissonrel} and \eqref{electricpoissonrel}, it
implies in
$
\nabla_i\left(\sqrt{Z\,}\nabla^i\varphi/{W}\right) = 0
$
where here
$
Z =  G_d- {W^\prime}^2  \, . \label{Zdef}
$
This equation is to be compared to its Newtonian analog and, in fact,
has the same form.  So it is possible to generalize the theorem
by De and Raychaudhuri \cite{de-raichaudhuri} to higher dimensions
(see also \cite{lemoszanchin1}).  According to such a theorem, in
order to have charged dust solutions satisfying Weyl hypothesis
without singularities, the quantity $Z$ must vanish. This implies
${W^\prime}^2 = G_d, $ as in the Newton-Coulomb with electric
matter theory, so that the result is the Majumdar-Papapetrou
relation, the same as in the relativistic vacuum case (see Eq.
(\ref{MPsansatzpotentialsrel})),
$
W = a_0 -\epsilon\sqrt{G_d}\, \varphi\, ,
\label{mprelationrel}
$
with $\epsilon = \pm 1,$ and $a_0$ being an integration constant.
Then, substituting $W$ from the latter equation into
Eq. (\ref{conserveq2}) gives $\rho_{\rm e} =
\epsilon\,\sqrt{G_d\,}\rho_{\rm m}$ as in the Newtonian case.  To sum
up, let us write the resulting equations for the important functions
$W$, $\varphi$, $\rho_{\rm m}$, and $\rho_{\rm e}$. In order to get a
field equation similar to Poisson equation \eqref{poissonfinal}, it is
convenient to introduce a new potential $U$ such that
\beq
U = \frac1W\,.
\label{u}
\eeq
The relevant equations are then
\beqa
& &\nabla^2 U=-S_{d-2}\,G_d \,U^{d-1\over d-3}\,\,\rho_{\rm m} ,
\label{poissonfinalrel} \\
& &\varphi=\epsilon \frac{1}{\sqrt{G_d\,}}\left(1- \frac 1 U \right) \,,
\label{mppotentialsfinalrel}\\
& &\rho_{\rm e} = \epsilon\,\sqrt{G_d\,}\rho_{\rm m}\, ,
 \label{mpchargefinalrel}
\eeqa
where an arbitrary constant in the potentials was adjusted to unity.
Some special solutions to these type of systems are going to be
analyzed in the next sections.  Eq. (\ref{mpchargefinalrel}) is the
Majumdar-Papapetrou condition.  Note, that these equations can be
compared to the Newton-Coulomb with an electrically charged
fluid case. In fact, taking the Newtonian limit in which $U\simeq 1
-V$, with $ |V| << 1$, one sees that Eqs.  \eqref{poissonfinalrel},
\eqref{mppotentialsfinalrel} and \eqref{mpchargefinalrel} reduce
exactly to Eqs. \eqref{poissonfinal}, \eqref{electricfinal} and
\eqref{chargefinal}, respectively.

\subsection{Spherical $d$ spacetime dimensional relativistic Bonnor star
solutions}
\label{bonnorrel}

\subsubsection{Equations in spherical coordinates}

In what follows we confine attention to spherically symmetric static
spacetimes and write the foregoing equations in isotropic and
Schwarzschild spherical coordinates.

\vskip 0.1cm
\noindent Equations in isotropic coordinates:
The starting point is the metric \eqref{metric},
which with Eq. (\ref{u}) now reads
\begin{eqnarray}
ds^2 = -U^{-2} dt^2 + {U^{2\over d-3}}\left(dR^2
+ R^2 d\Omega_{d-2}^2 \right) \, , \label{sphmetric}
\end{eqnarray}
with $d\Omega_{d-2}^2$ being the metric on the unit
$(d-2)$-dimensional sphere $S^{d-2}$.
$U$ is now a function of the radial coordinate $R$
only, and it obeys
\begin{equation}
\frac{d}{dR}\left(R^{d-2}\frac{dU}{dR}\right) = -S_{d-2}\,G_d\,\rho_{\rm
m}
R^{d-2}
U^{d-1\over d-3}\, ,
\label{harmonicpoisson}
\end{equation}
which is obtained from Eq. \eqref{poissonfinalrel}.
The matter and charge densities are also
functions of $R$ only, $\rho_{\rm m} =\rho_{\rm m}(R)$ and $\rho_{\rm
e} =\rho_{\rm e}(R)$, and they are related to each other through Eq.
(\ref{mpchargefinalrel}).  From Eq. (\ref{harmonicpoisson}), and
in analogy with the Newtonian theory,
define the mass function $m(R)$ and the charge function $q(R)$, (see
other mass function definitions in Appendix \ref{appendixmass}), as
\beqa
m(R) &=& S_{d-2}\int_0 ^R \rho_{\rm m}(R)\, U(R)^\frac{d-1}{d-3} R^{d-2}
dR\,,
\label{harmonicmasseq}\\
q(R) &=& S_{d-2} \int_0^R \rho_{\rm e}(R) \, U(R)^\frac{d-1}{d-3}  R^{d-2}
dR\,.
\label{harmonicchargeeq}
\eeqa
Eqs. \eqref{poissonfinalrel}-\eqref{mpchargefinalrel} may then be written
as
\beqa
\frac{dU(R)}{dR} & =& -G_d \frac {m(R)}{R^{d-2}}\, ,
\label{poissonrelisotropic}\\
\frac{d\varphi(R)}{dR} & =& -U(R)^{-2}\,\frac{q(R)}{R^{d-2}} \, ,
\quad {\rm or} \quad
\varphi(R)=
\frac{\epsilon}{\sqrt{G_d\,}}\left(1-\,\frac{1}{U(R)}\right)\,,
\label{poissonelectricrelisotropic}\\
q(R)&=& \epsilon\,\sqrt{G_d\,} m(R)\, ,
\label{mpchargerelisotropic}
\eeqa
where arbitrary constants in the potentials were set to one.
In addition, terms of the form ${\rm const}/R^{d-2}$ in Eqs.
\eqref{poissonrelisotropic}
and \eqref{poissonelectricrelisotropic}
were not written explicitly since they are implicitly
absorbed in those equations, and moreover they
should be put to zero as the fields shall be
regular functions of the radial coordinate $R$.
Eqs. (\ref{poissonrelisotropic})-(\ref{mpchargerelisotropic})
can then be compared to the Newton-Coulomb
with an electrically charged fluid case. In fact,
taking the Newtonian limit in which $U\simeq 1- V$, with $
|V| << 1$, and $R \simeq r$, one sees that Eqs.
\eqref{poissonrelisotropic}-\eqref{mpchargerelisotropic} reduce exactly
to
Eqs. \eqref{gravityeq}-\eqref{mq}, respectively.
For future reference we write here the Kretschmann ($\cal K$) and Ricci
($\cal R$) scalars for the metric
(\ref{sphmetric}):
\begin{eqnarray}
{\cal K}&=& {d-1\over d-3}\,{4{U^\pprime}^2\over
U^{2(d-1)\over d-3} }+
\left(8 + {(3d-8)(4d-11)\over (d-3)^3}\right){2{U^\prime}^4
\over U^{4(d-2)\over d-3}}
-{(d-2) (2d-5)\over (d-3)^2}\,{4{U^\prime}^2U^\pprime
\over U^{(3d-5)\over d-3}}\nonumber \\
&& + {8S_{d-2} \over d-3}\,G_d\,{\rho_{\rm m}\over U^{d-1\over d-2} }
    \left(U^\pprime - {U'^2\over U} +
\frac{S_{d-2}}{2} G_d {d-2\over d-3}  \rho_{\rm m} U^{d-1\over
d-3}\right)
\,
,
\label{kretsch}\\
{\cal R}&=& {2S_{d-2} \over d-3}\,G_d\,\rho_{\rm m}-{d-4\over d-3}
{{U^\prime}^2
\over U^{2(d-2)\over d-3}}\, ,
\label{riccisc}
\end{eqnarray}
where the prime stands for the derivative with respect to $R$.  {From}
this it is seen that spacetime singularities occur at points where
$U=0$, as long as the derivatives of $U$ do not vanish at the same
points as $U$ does.  Although the field equations are easily written
and solved by working in harmonic coordinates, the physical
interpretation of the solutions is clearer if one uses Schwarzschild
coordinates.

\vskip 0.1cm
\noindent Equations in Schwarzschild coordinates:
In Schwarzschild coordinates the
line element reads
\begin{equation}
ds^2= -B^2 dt^2 + A^2\, dr^2 + r^2\, d\Omega_{d-2}^2\, ,
\label{schwmetric}
\end{equation}
where $B=B(r)$ and $A=A(r)$, $r$ being the Schwarzschildean radial
coordinate.
By comparing
Eq. \eqref{sphmetric} to Eq. \eqref{schwmetric}, we see
that the radial coordinates in the two systems are related by
\begin{equation}
r^{d-3}=U\,R^{d-3} \, ,\label{radialcoords}
\end{equation}
and that the metric potentials are related by
\begin{equation}
      B = {1 \over U}\, , \label{gttcoeff}
\end{equation}
and
\begin{equation}
     {A } = 1 - {1\over d-3\,}{r \over U}\, {dU \over dr} \, .
\label{grrcoeff}
\end{equation}
Eq.~(\ref{radialcoords}) gives $r$ as a function of $R$. Although this
implicitly determines $R$ as a function of $r$, it is only in special
cases that this relation can be worked out explicitly.
For the sake of completeness, we present here the Schwarzschild
coordinate
form of the field equations.  With the metric in the form of
Eq.~(\ref{schwmetric}), Eq. (\ref{poissonfinalrel}) turns into
\begin{equation}
\frac{1}{A}\frac{d}{dr}\left(
{r^{d-2}}{1\over AB}\frac{dB}{ dr}\right)
= S_{d-2}\,G_d\, r^{d-2}\rho_{m}\,.
\label{schwarzpoisson}
\end{equation}
This is, in fact, the equation for the potential $B$, since $A$ is not
independent of $B$. Namely, Eqs. (\ref{gttcoeff}) and (\ref{grrcoeff})
give
\beq
{A} = 1 + {1\over d-3\,}{r \over B}\, {dB \over dr} \, ,
\label{AtoBrelation}
\eeq
which is a consequence of the Majumdar-Papapetrou condition in a
fluid with vanishing stresses.
At last, the electric functions are expressed in Schwarzschild
coordinates.
No effort is needed to obtain the electric charge density since it is
proportional the the mass density. The electric potential $\varphi(r)$
comes after Eqs. \eqref{mppotentialsfinalrel} and \eqref{gttcoeff}, i.e.,
\beq
\varphi = \frac{\epsilon}{\sqrt{G_d\,}} \left(1- B \right)\, ,
\label{mppotentialSchw}
\eeq
where, as usual, the arbitrary constant was  set to unity.
Now, defining ${\cal M}(r)$ and ${\cal Q}(r)$,
(see for comparison other mass definitions in Appendix
\ref{appendixmass}), as
\beqa
{\cal M}(r) &=& S_{d-2}\int_0 ^r \rho_{m}(r)\, A(r) r^{d-2} dr\,
,\label{schwarzmasseq}\\
{\cal Q}(r) &=& S_{d-2} \int_0^r \rho_{\rm e}(r) \, A(r) r^{d-2} dr\, ,
\label{schwarzchargeeq}
\eeqa
Eqs.\eqref{poissonfinalrel}-\eqref{mpchargefinalrel} may be written as
\begin{eqnarray}
\frac{dB(r)}{dr} & =& G_d A(r)\, B(r)\frac {{\cal M}(r)}{r^{d-2}}\, , \\
\frac{d\varphi(r)}{dr} & =& - A(r)\,B(r)       \frac{{\cal Q}(r)}{r^{d-2}}
\,,\quad {\rm or } \quad
\varphi(r) = \frac{\epsilon}{\sqrt{G_d\,}} \left(1- B(r) \right)\, ,
\\
{\cal Q}(r)&=& \epsilon\,\sqrt{G_d\,} {\cal M}(r)\,,
\end{eqnarray}
and $A(r)$ is given in terms of $B(r)$ by Eq. (\ref{AtoBrelation}).
These are the fundamental equations in Schwarzschild coordinates. The
Newtonian
limit is obtained by noticing that for weak gravity fields one has that
the metric functions $B(r)$ and $A(r)$ are close to unity,
$B(r) = 1+\delta B(r)$, and $A(r)=1+\delta A(r)$,
with $\delta$ indicating small
quantities. Hence, to the first order approximation, the
above equations reduce respectively to Eqs. \eqref{gravityeq}-\eqref{mq}.

\subsubsection{Solutions}
\label{ddmpsphericalsolutions}

{\it\small (a) Electrovacuum solutions in $d$ spacetime dimensions}
\vskip 0.1cm

As a first example and to set up notation let us report here on the case
of $d$-dimensional vacuum Majumdar-Papapetrou solutions. These are
nothing
but the extreme Reissner-Nordstr\"om spacetimes generalized to higher
dimensions that were first studied in Ref. \cite{tangherlini}.
 The general solution of
Eq.~ \eqref{poissonrelisotropic}
in vacuum is usually written in the form
\beqa
U &=& 1 + \frac{G_d} {d-3}\frac{M}{R^{d-3}}\, ,\label{Uisotropicvacuum}\\
M & =& {\rm constant}\, ,\label{Misotropicvacuum}
\eeqa
where $M$ is an integration constant equal to the
total mass of the source. The electric potential follows from Eq.
\eqref{poissonelectricrelisotropic}, $\phi = \epsilon\left(1- 1/U\right)
/\sqrt{G_d}$,
and the electric charge is related to the total mass of the source by
$Q = \epsilon\,\sqrt{G_d\,} M$, as required by the Majumdar-Papapetrou
condition and in agreement with Eq. \eqref{mpchargerelisotropic}.
The corresponding spacetime metric is
\beq
ds^2 = -\left(1+ \frac{G_d}{d-3} { M\over R^{d-3}}\right)^{-2} dt^2 +
\left(1+ \frac{G_d}{d-3}
{M\over R^{d-3}}\right)^{2\over d-3} \left(dR^2 +
R^2d\Omega_{d-2}^2\right)\, .
\label{ddRN}
\eeq
Using
Eqs.~\eqref{radialcoords}-\eqref{grrcoeff}
we find the relation between $r$ and $R$, given by
\begin{equation}
r^{d-3} = R^{d-3} + \frac{G_d}{d-3}{M} \, . \label{ddRN_Rtor}
\end{equation}
One also finds that
$B={1\over A} = \left({R^{d-3}\over R^{d-3} +\frac{G_d}{d-3}{M}}\right) =
          \left(1 - \frac{G_d}{d-3} {M \over r^{d-3}} \right)  \,
$
which leads to the metric for an extreme Reissner-Nordstr\"om
black hole with  mass and charge equal to $M$, and holds
for all $d\geq 4$,
\beq
ds^2= -\left(1- \frac{G_d}{d-3}{M\over r^{d-3}}\right)^2\, dt^2
+{dr^2\over  \left(1- \frac{G_d}{d-3}{M\over r^{d-3}}\right)^2}
+r^2d\Omega^2_{d-2}\, .
\eeq
The coordinate $r$ can be extended up to $r=0$, which is  in fact a
singularity. This is seen from the Ricci and Kretschmann scalars which
are, respectively,
\beqa
 {\cal K} &=& {4} \frac{(d-1)(d-2)^2}{d-3}\,\frac{G_d^2M^2}{r^{2(d-1)}}+
{2} \left(8 + {(3d-8)(4d-11)\over (d-3)^3}\right)
\frac{G_d^4 M^4}{r^{4(d-2)}} \nonumber\\
& & -{4 } \frac{(2d-5)(d-2)^2}{(d-3)^2}\frac{G_d^3M^3}{r^{(3d-5)}}\,,\\
{\cal R} &=&-\frac{d-4}{d-3}\,{G_d^2M^2\over r^{2(d-2)}}\, ,
\eeqa
where we used Eqs. \eqref{kretsch}, \eqref{riccisc},
\eqref{radialcoords} and (\ref{ddRN_Rtor}).
The region of the spacetime which in Schwarzschild coordinates
corresponds to $0 < r\leq (G_d\,M/(d-3))^{1/(d-3)}\,$ is not
covered by the isotropic coordinates. The maximal analytical
extension of these vacuum solutions representing extremal black holes
can then be found following the usual methods.

\vskip 0.3cm
{\it\small (b) Relativistic Bonnor star solutions in $d$ spacetime
dimensions}
\label{ddbonnorstars}

Interesting exact solutions in the context of Majumdar-Papapetrou
relativistic systems are the Bonnor stars, see now specifically
\cite{bonnor72,bonnor75,bonnor99}, which are spherically symmetric
distributions of a charged dust fluid satisfying the Einstein-Maxwell
with matter
equations in four-dimensional spacetimes.  The $d$-dimensional version
of such kind of stars are solution to Eq. \eqref{harmonicmasseq}, or
Eq. \eqref{schwarzpoisson}, with appropriate boundary and matching
conditions. We look for solutions using
the equations in harmonic coordinates, and then do the analysis in
Schwarzschild coordinates. In order to find solutions to
Eq. \eqref{harmonicmasseq}, a first, possible, procedure
is to provide the mass
density as a function of the radial coordinate, $\rho_{\rm m}=
\rho_{\rm m} (R)$. This is the procedure usually adopted, because it
furnishes by construction physically acceptable mass distribution for
the star.  In the present case, however, such a strategy is not
advisable because it results in a second order non-linear
differential equation for $U(R)$, whose solutions can be found just
after fixing the number of dimensions of the spacetime. A second
procedure, of no interest in the Newtonian case, but valuable here,
is to choose the energy density profile in such a way to
transform Eq.  \eqref{harmonicmasseq} into an equation whose
solutions are known, such as the case of sine-Gordon equation used in
Ref. \cite{varela03}, or transforming it into a linear equation, so
that one can use the well known methods to solve ordinary linear
second order differential equations to find solutions. A third
alternative procedure is to fix a priori the metric potential $U = U(R)$,
and then determining the other physical quantities that follow from
it. This is the strategy we follow here, it allows to
write the solutions in closed form, and it is the same
strategy as the one opted for in the Newtonian Bonnor stars
studied above.

\vskip 0.3cm
{\it\small (i) Solutions with smooth boundary conditions and some special
solutions}


First we make the analysis in isotropic coordinates. We consider the
general relativistic analog of the one studied in Sec.
\ref{newtoniansolutions}(b) (see also \cite{bonnor72,bonnor75,bonnor99}).
We then choose
\begin{equation}
U = \left\{\begin{aligned}
& U_{\rm i}=c_0+c_1 R^\alpha+c_2R^\beta\,, & R\leq R_0\,,\\
&  U_{\rm e}=1 + \frac{1}{d-3}\frac{G_dM }{R^{d-3}} \, ,& R >R_0\, .
\end{aligned}\right. \label{harmonicpotential0}
\end{equation}
where $\alpha$ and $\beta$
are real numbers and $R_0$ shall be identified as the
surface of the star. The arbitrary constants $c_0$, $c_1$, and $c_2$ are
fixed in such a way to guarantee the matching conditions at the surface
of the star, $R=R_0$.
Bonnor \cite{bonnor72,bonnor75,bonnor99} imposed $U$ to be a $C^1$
function
and the energy density to be a step function at the boundary.
In this case one can verify that the constants are given by
$c_0=1+\frac{G_d}{d-3}\frac{M}{R_0^{d-3}}\,\left(\frac{\beta+d-3}{\beta}
\right)
+c_1\,\frac{\alpha-\beta}{\beta}\,R_0^\alpha$, $c_1$ one can take
as arbitrary, and
$c_2=-\frac{G_d}{\beta}\frac{M}{R_0^{\beta+d-3}}
-\frac{\alpha}{\beta}\,c_1\,R_0^{\alpha-\beta}$. To reproduce
Bonnor's choice for $U$ \cite{bonnor99} one puts $d=4$,
$c_1=0$ and $\beta=n$ (where $n$ was the letter chosen for the
exponent in \cite{bonnor99}).
Of course, if one wishes, one can choose $U$ to
be of any degree of differentiability at the boundary.
Since it is interesting to test whether this choice of differentiability
has any important influence on the properties of the star one
can, still in the spirit of Bonnor, go a step
further and instead of choosing $U$ as a $C^1$ function,
impose $U$ to be a $C^2$ function of $R$. As a bonus, one gets
in addition, that
the energy density is a $C^0$ function, i.e., continuous at the
boundary $R_0$, indeed zero, which is more in accord with the
usual properties of stars. For a $C^2$ choice for $U$
there are no free constants
and one finds,
\beqa
&& c_0 =1+\frac{1}{d-3}{G_dM\over R_0^{d-3}}\left[1+{d-3\over
\beta-\alpha}\,
\left({\beta+d-3\over \alpha}- {\alpha+d-3\over\beta}\right)
          \right]\, ,\label{c0}\\
&& c_1 =-{(\beta+d-3)\over \alpha(\beta-\alpha)}
{G_dM\over R_0^{\alpha+d-3}}\, , \label{c1}\\
&& c_2 = - {(\alpha+d-3)\over \beta(\alpha-\beta)}
{G_dM\over R_0^{\beta+d-3}}\, .\label{c2}
\eeqa
It then follows the potentials $U_{\rm i}$ and
$U_{\rm e}$ are
\begin{eqnarray}
U = \left\{\begin{aligned}
U_{\rm i}= &1+\frac{G_d}{d-3}\frac{ M}{R_0^{d-3}}\left(\!
1 + \frac{(d-3)(\beta+d-3)}{\alpha(\beta-\alpha)}
\left[1- \left(\frac{R}{R_0}\right)^{\!\alpha}\right]\right. \\
&\left.-\frac{(d-3)(\alpha+d-3)}{\beta(\beta-\alpha)}\left[1-
\left(\frac{R}{R_0}\right)^{\!\beta}\right] \right),
& R\leq R_0\, , \\
U_{\rm e}=&1 + \frac{1}{d-3}\frac{G_dM }{R^{d-3}} \, ,& R >R_0\, .
\end{aligned}\right. \label{harmonicpotential}
\end{eqnarray}
Eq. \eqref{poissonfinalrel} then gives the mass density
\begin{equation}
\rho_{\rm m} =\!\left\{\!\begin{aligned}
& \frac{(\alpha+d-3)(\beta+d-3)}{(d-3)(\beta-\alpha)}
\frac {M} {S_{d-2}\, R_0^{d-1}}\!\left[
\left(\frac{R}{R_0}\right)^{\alpha-2}-\left(\frac{R}{R_0}\right)^{\beta-2
}
\right]\frac{1}{U^{d-1\over d-3}}\, ,& R\leq R_0\,, \\
  & 0 \, , & R>R_0  \, .
\end{aligned} \right.
\label{harmonicmass}
\end{equation}
In the region outside the mass
distribution, the solution takes the extreme Reissner-Nordstr\"om form
(\ref{ddRN}), as expected. Since $U$ is a $C^2$ function, the spacetime
metric satisfies the Israel matching conditions at $R=R_0$.
In order that $\rho_{\rm m}$ be a well defined function
and everywhere non-negative
we must have $\alpha, \beta \geq 2$.
The quantity $M/(S_{d-2}\, {R_0}^{d-1})$ appears naturally
with units of mass density.
Note that the electric potential $\varphi$ and the the electric
density $\rho_{\rm e}$ can be found directly from Eqs.
(\ref{poissonelectricrelisotropic}) and (\ref{mpchargerelisotropic}).
The function $\varphi$ is a continuous $C^2$ function through the
surface of the star, which means the field strength is $C^1$ and the
charge density is $C^0$.  Moreover, using Eqs. (\ref{harmonicmasseq})
and (\ref{harmonicmass}) one finds that indeed $M=m(R_0)$, making the
whole procedure a consistent one.  This Bonnor star solution looks
like the Newtonian star studied in Sec.  \ref{ndnewtonianstars}. In
fact, the resulting mass density, Eq. \eqref{harmonicmass}, 
resembles the function given by Eqs. \eqref{sphericalmass}.

Second, we make the analysis in Schwarzschild coordinates.
Schwarzschild coordinates are interesting to analyze the physical
properties of the spherical solutions found above.
Eqs. \eqref{radialcoords} and \eqref{harmonicpotential} establish the
relation between the harmonic radial coordinate $R$ and the
Schwarzschild radial coordinate $r$
\beq
r^{d-3}=\left\{\begin{aligned}
&  c_0 R^{d-3}+c_1 R^{\alpha+d-3}+c_2 R^{\beta+d-3}\,, & R\leq R_0\,,\\
& R^{d-3} +\frac{1}{d-3}\, G_dM\, ,&    R > R_0\, .
  \end{aligned}\right. \label{relation_rtoR}
\eeq
These relations furnish $R$ as a function of $r$, $R = f(r)$, which is
in fact defined by two functions. Let us call them respectively
 $f_{\rm i}(r)$, for the internal
region, and $f_{\rm e}(r)$, for the external region. The surface of the
star,
defined by $R=R_0$, is obtained in terms of the Schwarzschild
coordinates,
by imposing the continuity of the function $r(R)$ through
such a surface, i.e.,
\beq
{r_0}^{d-3}={R_0}^{d-3}U_{\rm i}(R_0)={R_0}^{d-3}U_{\rm
e}(R_0)={R_0}^{d-3}+\frac{1}{d-3}\, G_dM \,,
\eeq
where $U_{\rm i}$ and $U_{\rm e}$ are defined by Eq.
\eqref{harmonicpotential}.
The aim now is to find the metric potentials $B$ and $A$ as functions
of $r$.  In order to do that one needs to find the functions $f_{\rm
i}(r)$ and $f_{\rm e}(r)$, which is done by solving
Eqs. \eqref{relation_rtoR} for $R$.
For $r\leq r_0$ one has
\beqa
&&B_{\rm i}(r)=\frac{1}{U_{\rm i}(r)}=\left(c_0+c_1f_{\rm i}^\alpha
+c_2f_{\rm i}^{\beta}\right)^{-1}\,,\\
&&A_{\rm i}(r) =1+{r\over f_{\rm i}}\,{df_{\rm i}\over dr}
\left(\alpha c_1 f_{\rm i}^\alpha
  +\beta c_2f_{\rm i}^{\beta}\right)
\left(c_0 +c_1 f_{\rm i}^{\alpha}+ c_2 f_{\rm i}^\beta\right)^{-1}\, ,
\eeqa
with $f_{\rm i}= f_{\rm i}(r)$ being a suitable solution of the following
algebraic equation
\beq
c_2\,{f_{\rm i}}^{\beta+d-3}
+c_1 f_{\rm i}^{\alpha+d-3}+c_0 f_{\rm i}^{d-3}-r^{d-3}=0\, .
\label{relation_Rtor}
\eeq
The constants $c_0$, $c_1$ and $c_2$ are now to be
written in terms of $r_0$, instead of in terms of $R_0$.  The
corresponding expressions are obtained by substituting $R_0=\!\!
({r_0}^{d-3} \!-G_dM/(d-3))^{1/(d-3)}\,$ into Eqs.
\eqref{c0}--\eqref{c2}.
For $r\geq r_0$ one has \beq
B_{\rm e}(r)= \frac{1}{A_{\rm e}(r)} =
1- \frac{1}{d-3}{G_dM\over r^{d-3}}  \,.
\eeq


Third, we find some special solutions with a simple algebraic structure.
Generally, the only way of finding the solutions to Eq.
\eqref{relation_Rtor} is
by specifying the values of the parameters $\alpha$ and $\beta$ and
the number of spacetime
dimensions $d$. Even in that case, in general, only numerical solutions
are possible to find and we do not perform such an analysis here.
There are, however, some special values of $\alpha$ and $\beta$ for
which Eq. \eqref{relation_Rtor} can be solved exactly for $f_{\rm i}(r)$.
Thus, in order to investigate some more properties of
$d$-dimensional Bonnor stars, we consider a
particular case that can be dealt with
algebraically.  For instance, one  may choose
\beq
\beta = \frac32\,\alpha= 3(d-3)\, ,
\label{alphabetarelation}
\eeq
so that one finds a fourth degree polynomial equation to solve
for $R^{d-3}$:
\beq
c_2\left(R^{d-3}\right)^4+c_1\left(R^{d-3}\right)^3+
c_0 R^{d-3} - r^{d-3} =0 \, , \label{tosolveforR}
\eeq
where now the coefficients $c_0$, $c_1$ and $c_3$ are simplified to
\beq
c_0 = 1+\frac{2G_d}{d-3}\frac{M}{{R_0}^{d-3}}\, , \qquad
c_1 = -\frac{2 G_d}{d-3}\frac{M}{{R_0}^{3(d-3)}} \, ,\qquad
c_2=\frac{G_d}{d-3}\frac{M}{{R_0}^{4(d-3)}}\, .\label{c012}
\eeq
This polynomial equation can be solved in terms of radicals,
and the physical quantities can then be expressed explicitly in terms
of the coordinate $r$.
In order to condense expressions, we first define the parameter $a$
by
\begin{equation}
 a = \frac{G_d}{d-3}\frac {M}{{r_0}^{d-3}}\, ,\label{masstoradius}\\
\end{equation}
with $0\leq a\leq 1$.
As in the case of Newtonian stars (see Eq.\eqref{parameter-a}) this
parameter
measures how compact is the star and it is useful to parameterize the
numerical solutions.
Further, we define
\beqa \begin{aligned}
& b(r) =\frac{1}{16}\left(1+\frac{1}{a}\right)^2
-\frac{1}{4a}\,\left(\frac{r} {r_0}\right)^{d-3}\, ,
& c(r)= \frac{1}{6}+\frac{1}{6a}-
\frac{1}{3a}\left(\frac{r}{r_0}\right)^{d-3}\, ,\\
& e(r)=\left(b(r) +
\sqrt{ \left[b(r)\right]^2 - \left[c(r)\right]^3\,}\right)^{1/3}\, ,
\label{def1}
& s(r) = \sqrt{1 +2\,e(r)
+2\frac{c(r)}{e(r)}\, }\, , \end{aligned}
\eeqa
where we have used the relation ${R_0}^{d-3}=
{r_0}^{d-3} -\frac{G_d}{d-3} M$.
Then, the solution for $R(r)$ is
\beq  \label{solRtor}
{R(r)}^{d-3}= \left\{ \begin{aligned}
&\!\!\left(\frac{1}{2} - \frac{s(r)}
{2} + \frac{1}{2}\sqrt{2 + 2s(r) +2 \frac{c(r)}{\,e(r)}-
\frac{2}{a\,s(r)}\,} \right)\!\! \left({r_0}^{d-3} -\frac{G_d}{d-3}M
\right)\,,  & r\leq r_0\,,\\
&\displaystyle{r^{d-3} - \frac{G_d}{d-3}M\,,} & r>r_0\, . \end{aligned}
\right.
\eeq

Fourth, the relevant functions
$B(r)$, $A(r)$, $\rho_{\rm m}(r)$, $\varphi(r)$, and $\rho_{\rm e}(r)$,
given in terms of the Schwarzschild coordinates follow from the
above relations. They are dependent on the variable $r$,
and also depend on two other arbitrary parameters, the mass and the
radius of the star, $M$ and $r_0$, respectively. Instead of writing
the explicit form of such functions, which are cumbersome, it is more
convenient to plot them for several choices of parameters. In the
calculations we normalized the coordinate $r$ to the mass
parameter $\mu= ({G_d\,}M/({d-3}))^{1/(d-3)}$ which was kept fixed. In
fact, the important parameter to this end is the mass to radius ratio
$a$, given by Eq. (\ref{masstoradius}), which measures how
relativistic is the system. Here we have the constraint $0<a<1$, and
for small $a$ the system is Newtonian, while for $a$ close to unity it
is fully relativistic.
\begin{figure}\begin{center}
\includegraphics[scale=.8]{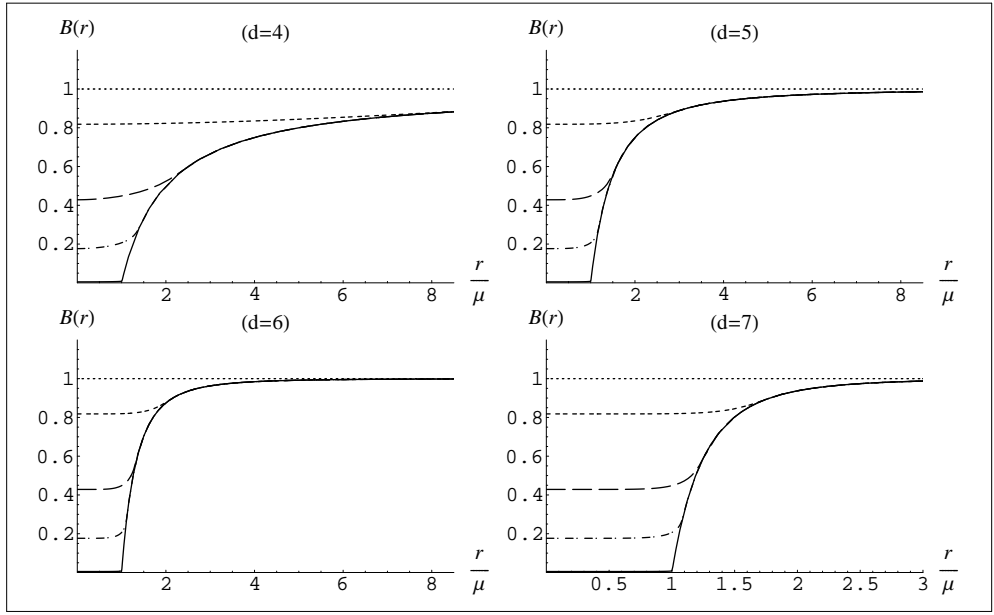}
\caption{The metric potential $B(r)$ as a function of
$r/\mu$, where $\mu\equiv (G_dM/(d-3))^{1/(d-3)}$, for $d=4,5,6,7$, and
for four values of $a$ in each graph (from top to bottom:
$a=0.1$, $a=0.4$, $a=0.7$ and $a=1$).}
\label{figure_gtt}\end{center}
\end{figure}
\begin{figure}
\begin{center}
\includegraphics[scale=.88]{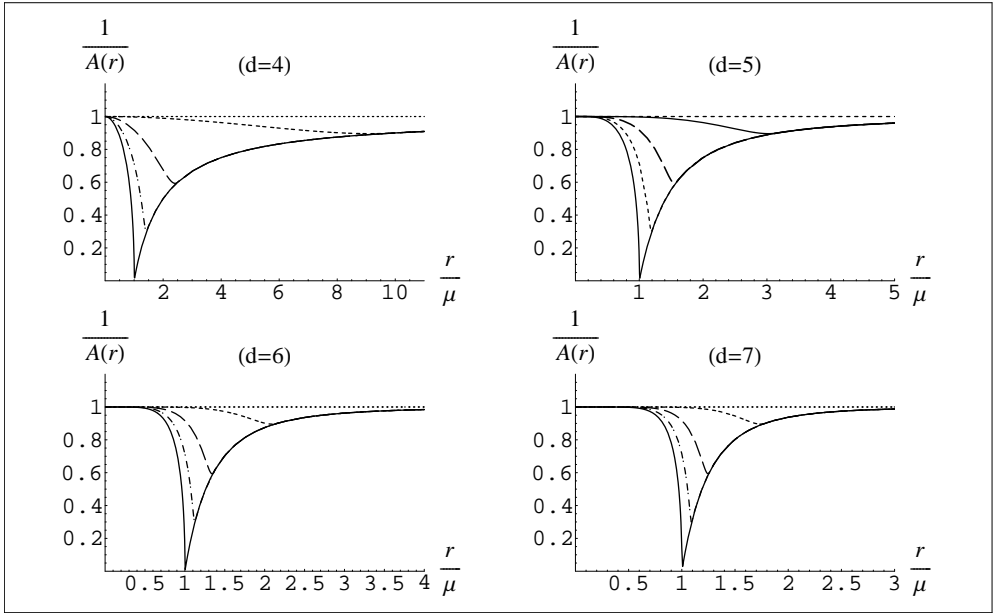}
\caption{The metric potential $1/A(r)$ as a function of
$r/\mu$, for $d=4,5,6,7$, and
for four values of $a$ in each graph (from top to bottom:
$a=0.1$, $a=0.4$, $a=0.7$ and $a=1$).}
\label{figure_grr}
\end{center}
\end{figure}
The function $B(r)$:
The simplest function to be found in Schwarzschild coordinates is the
metric potential $B(r)$, which is immediately obtained through the
relation $B(r) = 1/U(r)$. Fig. \ref{figure_gtt} shows $B(r)$ as
function of $r/\mu$ in $d=4$, $5$, $6$, $7$, as indicated. It is also
seen in that figure the behavior of $B(r,a)$ as a function of $a$, for
different values of the parameter $a$, as shown by the four curves in
each graph. All the interior functions $B_{\rm i}(r,a)$ match the
exterior extreme Reissner-Nordstr\"om solution $B_{\rm e}(r)= 1
-(\mu/r)^{d-3}$, each one at a different value of $r_0$. The
reason for that is because the change of $a$ is made by keeping the
mass of the star fixed, while $r_0$ varies accordingly.  Notice also
that for $a\rightarrow 1$ the function $B_{\rm i}(r)$ approaches zero
in the whole region interior to $r=r_0$, meaning that the redshift
with respect to infinity is infinite.  For the extreme value ($a=1$)
the mass and the charge of the charged star are concentrated inside a
quasihorizon at $r=r_0$.  In this limit, the spacetime solution is a
quasi-black hole, similar to what was found for four-dimensional
spacetimes (see \cite{kleberlemoszanchin}-\cite{lemoszaslavskii2}).
There are no singularities inside $r_0$, the curvature is finite, so
are the mass and charge densities of the charged dust (see also item
(ii) below).  It can also be seen the Newtonian limit of the solution
by comparing the curves for the smaller values of $a$ in
Fig. \ref{figure_gtt} with the corresponding curves for the Newtonian
potential, Fig.  \ref{figure_VN} (see also item (iii) below).
\begin{figure}
\begin{center}
\includegraphics[scale=.8]{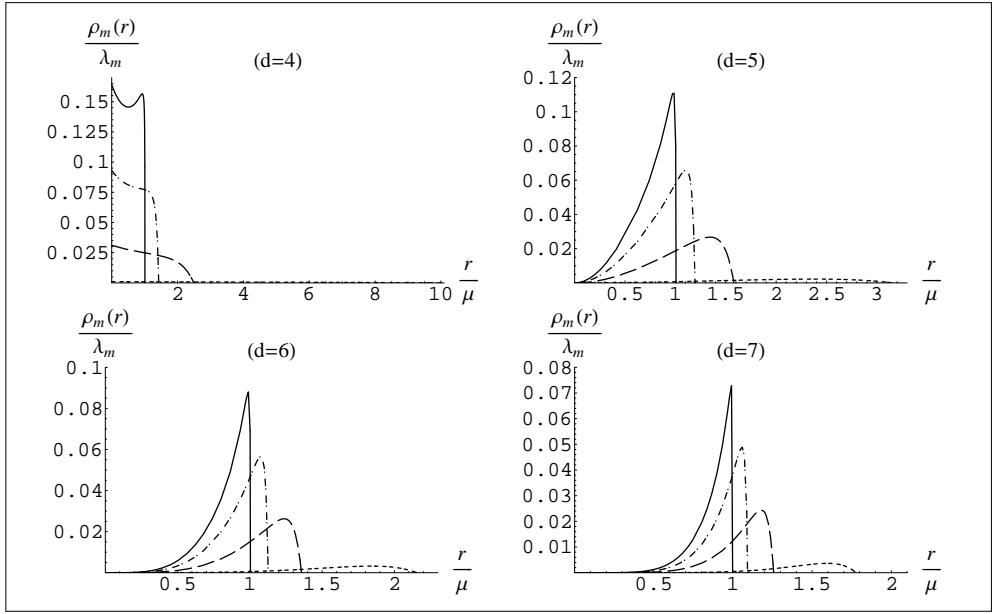}
\caption{The normalized relativistic mass density $\rho_{\rm
m}(r)/\lambda_{\rm m}$ as a function of $r/\mu$, where $\lambda_{\rm
m}=\frac{12}{d-1}\bar\rho_{\rm m}$, $\bar\rho_{\rm m}$ being a kind of
average density (see text), and $\mu\equiv (G_dM/(d-3))^{1/(d-3)}$,
for the cases $d=4$, $5$, $6$, $7$ (as indicated), and with $a=1$
(upper curve), $a=0.7$ (dot-dashed curve), $a=0.4$ (dashed curve), and
$a=0.1$ (lowest curve) for each $d$. The normalized relativistic
mass density $\rho_{\rm m}(r)/\lambda_{\rm m}$ goes to zero at the
surface of the star, defining thus the radius $r_0$ in each plotted
case.}
\label{figure_rho} 
\end{center}
\end{figure}
The function $A(r)$: The behavior of the other metric potential $A(r)$
is seen in Fig.  \ref{figure_grr}, where we plot $1/A$ against $r/\mu$
for the same values of $d$ and $a$ as in Fig. \ref{figure_gtt}.  The
quasi-black hole formation is seen in this case as $1/A(r)$ going to
zero at $r=r_0$ for $a\rightarrow 1$. It appears in the figure as the
sharp elbow in the solid line (lowest) curve showed in the graph.
The exterior function is $ A_{\rm e}(r)= 1/B_{\rm e}(r)$, and all
the inner functions $A_{\rm i}(r,a)$ for different $a$ match $A_{\rm
e}(r)$ at a particular value of $r_0$. 
The function $\rho_{\rm m}(r)$: Another quantity of interest is the
mass density $\rho_{\rm m}(r)$.  In Fig. \ref{figure_rho} we plot the
normalized mass density $\rho_{\rm m}(r)/{\lambda_{\rm m}}$ as a
function of the normalized radial coordinate $r/\mu$. Here
$\lambda_{\rm m}$ is defined as $\lambda_{\rm m}=
\frac{(\alpha+d-3)(\beta+d-3)}{(d-1)(d-3)(\beta-\alpha)}\bar\rho_{\rm
m}$, where $\bar\rho_{\rm m}$, a kind of average density, here is
given by $\bar\rho_{\rm m}={(d-1)M}/(S_{d-2}\, r_0^{d-1})$. For our
choice of parameters, see Eq. (\ref{alphabetarelation}), one has
$\lambda_{\rm m}=\frac{12}{d-1}\bar\rho_{\rm m}$.  We plot $\rho_{\rm
m}(r)/{\lambda_{\rm m}}$ against $r/\mu$ for the same values of $d$
and $a$ as in Figs. \ref{figure_gtt} and \ref{figure_grr}.  
Notice that for $d>4$ the general properties of this function do not
depend upon the specific value of $d$.
It is clearly seen that $\rho_{\rm m}(r)$ is
finite at $r=0$. In fact,
with our choice, $\rho_{\rm m}$ vanishes at $r=0$ for all
$d>4$. In addition it goes to zero at the surface of the star, 
defining thus the radius $r_0$ in each plotted case. 
Moreover, the mass density is everywhere well defined even in
the quasi-black hole limit.  
The comparison to the Newtonian case can be done considering the
curves for small $a$ in Fig. \ref{figure_rho}, and comparing the
corresponding curves in Fig.  \ref{figure_rhoN}, (see below item
(iii)).  The functions $\varphi(r)$ and $\rho_{\rm e}(r)$: The other
two functions, the electric potential $\varphi(r)$ and the electric
charge density $\rho_{\rm e}(r)$, are so close related to the
respective gravitational quantities $B(r)$ and $\rho_{\rm m}(r)$, that
no plot need to be drawn for them. In fact, they are promptly obtained
from their relations to the functions studied above (see Eqs.
\eqref{mppotentialSchw} and \eqref{mpchargefinalrel}), namely,
$\varphi(r) =\epsilon (B(r)-1)/\sqrt{G_d\,},$ and $\rho_{\rm e}(r)=
\epsilon\sqrt{G_d\,}\rho_{\rm m}(r)\,.$

\vskip 0.3cm
{\it\small (ii) The quasi-black hole limit}
\label{quasi-black hole}

For the full relativistic limit, $a=1-\varepsilon$, with
$\varepsilon<<1$, it is clear from the previous plots that the
function $1/A(r)$ attains a minimum at $r/\mu=1+\varepsilon$, such
that $1/A(r)=\varepsilon$, where again, $\mu=
({G_d\,}M/({d-3}))^{1/(d-3)}$.
 Also, for such a small but non-zero $\varepsilon$ the configuration
is regular everywhere with a non-vanishing metric function
$B$. Moreover, in the limit $\varepsilon \rightarrow 0$ the interior
metric potential $B_{\rm i}$ obeys, $B_{\rm i}\rightarrow 0$, for all
$r/\mu\leq1$.  These three features define a quasi-black hole, see
\cite{lemoszaslavskii1,lemoszaslavskii2}.  These three features imply,
among other things, that (a) there are infinite redshift whole regions
rather than surfaces, (b) the object displays naked behavior, i.e.,
generation of infinite tidal forces in a freely falling frame, (c)
outer and inner regions become impenetrable and disjoint, and (d) for
external distant observers the spacetime is indistinguishable from
that of extremal black holes.  The quasi-black hole is on the verge of
forming an event horizon, but it never forms one, instead, a
quasihorizon appears.

It is of interest to see that in the quasi-black hole limit the metric
is well defined and everywhere regular.  We check this for the
interior.  One defines, from the isotropic radial
coordinate $R$, a new spatial coordinate $x$ by
\beq
x = \frac{R}{R_0}\, , \qquad \qquad 0\leq x\leq 1\, ,
\eeq
from which one sees that the surface of the star is now located at $x=1$.
Substituting this transformation into the interior metric functions and
choosing a new time coordinate $T$ according to
\beq
dT = \frac{(d-3) R_0^{d-3}}{G_d M }dt        \, ,
\eeq
the interior metric is now
\beq
ds^2 = -\tilde U^{-2} dT^2 +   \left(\frac{G_dM}{d-3}\right)^{2/(d-3)}
\tilde U ^{2/(d-3)}
\left(dx^2 + x^2 d\Omega_{d-2}^2\right) \, ,
\eeq
where
\beq
\tilde U = 1+ (d-3)\left[{\beta+d-3 \over \alpha(\beta-\alpha)}
\left(1-x^\alpha\right)+ {(\alpha+d-3)\over \beta(\alpha-\beta)}
\left(1 - x^\beta\right)\right] \,.
\eeq
This metric is regular throughout the interior region and also at the
surface of the star. Moreover, at $x=1$ one has $\tilde U=1$.  This
means that even in the quasi-black hole limit the surface of the star
is timelike for internal observers. On the other hand, one can verify
that, being the exterior metric the extremal Reissner-Nordstr\"om
metric, the quasi-black hole limit gives a null surface for external
observers.  There is thus a mismatch, implying in this case that the
interior and exterior regions are disjoint, as was fully analyzed in
\cite{lemoszaslavskii1,lemoszaslavskii2}.

\vskip 0.3cm
{\it\small (iii) The quasi-Newtonian limit:
the Newtonian Bonnor stars discussed previously}
\label{Newton_approx}

It is expected that in the weak field approximation a relativistic
Bonnor star reduces to a Newtonian Bonnor star. Here, we show that
indeed the relativistic star studied in this section, i.e.,
Sec. \ref{ddbonnorstars}, reduces to the Newtonian star studied in
Sec. \ref{ndnewtonianstars}.

In the relativistic theory two coordinate systems are involved in the
solutions, the isotropic and the Schwarzschild spherical coordinates.
Initially we show that to first order approximation in the weak field
limit the two coordinate systems are identical. In order to deal with
the issue we take the special case considered in paragraph (b)(i)
of subsection \ref{ddbonnorstars}.  The weak field limit inside the
spherical star corresponds to small values of the parameter $a=
G_dM/((d-3){r_0}^{d-3})$. Hence, considering the approximation of
Eq. \eqref{solRtor} up to the first order in $a$ it follows
\begin{equation}
\frac{R^{d-3}}{R_0^{d-3}}= \frac{r^{d-3}}{r_0^{d-3}}
\left\{1- \frac{G_d}{d-3}\,\frac{M}{r_0^{d-3}}\left[1-
2 \left(\frac{r^{d-3}}{r_0^{d-3}}\right)^2
 +\left(\frac{r^{d-3}}{r_0^{d-3}}\right)^3  \right]\right\} \,.
\end{equation}
At the lowest order approximation it results in
\begin{equation}
\frac{R^{d-3}}{R_0^{d-3}}=
\frac{r^{d-3}}{r_0^{d-3}}\, , \label{solRtor0}
\end{equation}
as expected.
Therefore, when comparing the first order approximation of
the relativistic solution to the Newtonian solution one may work with the
isotropic coordinates, identifying the radial coordinate $R$ with the
Newtonian radial coordinate $r$.

The next step is obtaining the potentials 
and the densities in the weak field
approximation and comparing them to the Newtonian case. In such a
limit one has the relation $U = 1-V$, where $V$ is the Newtonian
potential. Now using the relation \eqref{solRtor0} and Eq.
\eqref{harmonicpotential} one can write  $U_{\rm i}$ up to first order in
$M/{r_0}^{d-3}$,
\begin{eqnarray}
U_{\rm i} = 1+\frac{G_d}{d-3}\frac{ M}{r_0^{d-3}}\left(\!
1 + \frac{(d-3)(\beta+d-3)}{\alpha(\beta-\alpha)}
\left[1- \left(\frac{r}{r_0}\right)^{\!\alpha}\right] +\right.
\nonumber\\
\left.-\frac{(d-3)(\alpha+d-3)}{\beta(\beta-\alpha)}\left[1-
\left(\frac{r}{r_0}\right)^{\!\beta}\right] \right) \,.
\label{harmonicpotentialweak}
\end{eqnarray}
From this equation, and from the exterior solution $U_{\rm e}$, one then
finds
the potential in Newtonian approximation, $V= 1-U$, as
\begin{equation}
V  = \left\{ \begin{aligned}
V_{\rm i}=&-\frac{G_d}{d-3}\frac{M}{r_0^{d-3}}\left(1 +
\frac{(d-3)(\beta+d-3)}{\alpha\left(\beta-\alpha\right)}
\left[1- \left(\frac{r}{r_0}\right)^{\alpha}\right] +\right. \\
& \left.-\frac{(d-3)(\alpha+d-3)}{\beta\left(\beta-\alpha\right)}\left[1-
\left(\frac{r}{r_0}\right)^{\beta}\right]\right)\, ,
 & r\leq r_0\, ,\\
V_{\rm e} = &-\frac{G_d}{d-3}\frac{M}{ r^{d-3}} \, ,
 & r>r_0 \, .  \end{aligned} \right.
\label{sphericalpotentialr2}
\end{equation}
The resulting expression is to be compared to the gravitational potential
of the Newtonian star as given in Eq. \eqref{sphericalpotential}. The two
expression become identical if one identifies the gravitational constant
$G_d$, the radial coordinate $r$, and the mass of the star $M$ in both
equations.
We have already shown that, in the weak field approximation, it results
$ R = r + O(M/r_0^{d-3})$, and also $U(R)=U(r)= 1 + O(M/r_0^{d-3})$.
Therefore, substituting such results into Eq. \eqref{harmonicmass} we
find
the first order approximation for the relativistic mass density,
\begin{equation}
\rho_{\rm m} =\left\{\begin{aligned}
& \frac{(\alpha+d-3)(\beta+d-3)}{(d-3)(\beta-\alpha)}
\,\frac{M}{S_{d-2}\, r_0^{d-1}}\left[
\left(\frac{r}{r_0}\right)^{\alpha-2}-\left(\frac{r}{r_0}\right)^{\beta-2
}
\right] \, ,& r\leq r_0\,, \\
  & 0 \, , & r>r_0  \, .
\end{aligned} \right. \label{harmonicmassr2}
\end{equation}
In order for this result to be identical to Eq.  \eqref{sphericalmass}
the mass $M$ and the coordinate $r$ must be the same in both
equations.
It is then straightforward to show that the weak field limits of
other
relativistic quantities such as the metric functions $B(r)$ and
$A(r)$, the electric charge density, and electric potential all agree
with their Newtonian counterparts, as expected.

Notice that units have been normalized in such a way that the
gravitational coupling constant in Einstein equations equals to the
Newtonian gravitational coupling constant in Poisson equation (see
Appendix \ref{appendixgd}).  Furthermore, the mass densities carry
identical units and normalizations due to the similarity between
Poisson equation for Newtonian gravity, Eq. \eqref{gravitypoisson},
and the corresponding equation coming from Majumdar-Papapetrou
relativistic system, Eq. \eqref{poissonfinalrel}.

\section{Conclusions}
\label{conclusions}

We have studied $d$-dimensional Bonnor star solutions, spherical
distributions of extremal charged dust joined to extremal charged
vacua, both in Newtonian gravity and general relativity.  We have
found that the relativistic solutions present many interesting
properties such as forming an extreme $d$-dimensional quasi-black
hole, when the mass to radius ratio reaches a critical value.  We have
also found that the Newtonian solutions are limiting cases of the
relativistic ones. In this connection it is interesting to note that
the Bonnor star solutions in Majumdar-Papapetrou Newtonian gravity,
when contrasted to those Bonnor solutions in Majumdar-Papapetrou
general relativity, display clearly the departing of the high density
structures that may arise in the strong field regime of each theory,
mild singularities in one theory, quasi-black holes in the other.
Moreover, whereas there are no solutions for Newtonian stars
supported by degenerate pressure in higher dimensions, and so no
general relativistic solutions also, higher dimensional Bonnor stars,
supported by electric repulsion do indeed have solutions. This means
that the existence of stars in higher dimensions depends on the number
of dimensions itself, and on the underlying field content of those
stars, as expected.

\section*{Acknowledgments}
We thank conversations with Oleg Zaslavskii, Observat\'orio Nacional
of Rio de Janeiro for hospitality, and Centro Multidisciplinar de
Astrof\'\i sica at Instituto Superior T\'ecnico (CENTRA-IST) for a
grant and for hospitality while part of the present work has been
done.  JPSL thanks Funda\c{c}\~{a}o para a Ci\^{e}ncia e Tecnologia of
Portugal (FCT) through project POCI/FP/63943/2005 for financial
support.  VTZ thanks Funda\c c\~ao de Amparo \`a Pesquisa do Estado de
S\~ao Paulo (FAPESP) for financial help (No. 2007/04278-2), and
Conselho Nacional de Desenvolvimento Cient\'\i fico e Tecnol\'ogico of
Brazil (CNPq) for a fellowship.

\begin{appendix}

\section{Newton's gravitational
constant $G_d$ in $d$ spacetime dimensions}
\label{appendixgd}
\noindent Within Newtonian gravity, the Poisson equation
for the gravitational field is given by
\beq
\nabla^2 V =  k \rho_{\rm m}\,,
\eeq
where $k$ is a constant, related to Newton's gravitational
constant $G_d$ in $d$
spacetime dimensions, to be determined.
Integrating over the space volume ${\cal V}$ and using the Gauss theorem,
one obtains
\beq
\int_{\cal V} \nabla^2 V\, d^{d-1} x=\oint_{S_{d-2}}
\nabla_i V\,
{n^i}\, dS_{d-2}=k \int_{\cal V} \rho_{\rm m}\, d^{d-1}x = k\, M\,,
\label{2}
\eeq
where $S_{d-2}$ is the boundary surface surrounding the volume ${\cal
V}$,
and $n^i$ is the unit normal to the surface $S_{d-2}$.
Considering now spherical symmetry, i.e.,
\beq
\nabla_i V {n^i}= -g_r\,,
\eeq
where $g_r$ is defined to be the radial component of the
gravitational field, one finds
\beq
\oint_{S_{d-2}} \nabla_i V \,{n^i}
dS_{d-2} =  -g_r  S_{d-2}\, r^{d-2} \, .
\label{newold}
\eeq
Then (\ref{2}) and (\ref{newold}) yield
\beq
g_r = - {k\over S_{d-2}}{ M\over r^{d-2}}\, .
\eeq
The choice in  \cite{arkmed} for $k$ is given by
\beq
k = G_{d} S_{d-2}\,.
\label{k1}
\eeq
This is an interesting choice because it gives
\beq
g_r = - {G_{d} M\over r^{d-2}}\,,
\label{newtondef}
\eeq
i.e., a straight generalization of Newton's force law to $d$
spacetime dimensions, although it puts Einstein's equation
into a slightly awkward form,
\beq
G_{\mu\nu}= \frac{d-2}{d-3}S_{d-2}\,G_d\,T_\munu\,,
\label{einsteink1}
\eeq
where $G_{\mu\nu}$ is the Einstein tensor and
$T_\munu$ is the energy-momentum tensor.
The choice in \cite{myers} for $k$ is given by
\beq
k= 8\pi G_{d} {d-3\over d-2}\,.
\label{k2}
\eeq
This is also
interesting choice because although it gives
\beq
g_r = -{8\pi G_d\over S_{d-2}}{d-3\over d-2} { M\over r^{d-2}}\,,
\label{einsteindef}
\eeq
Einstein's equation are written as
\beq
G_{\mu\nu}=8\pi\,G_d\,T_{\mu\nu}\,,
\label{einsteink2}
\eeq
i.e., a  straight generalization of Einstein's equation
to $d$ spacetime dimensions.
Both definitions of $k$ give the
correct definition in four dimensions for $G_{d=4}=G_4\equiv G$.
In this paper we have opted for
the definition (\ref{k1}), which yields (\ref{newtondef})
and (\ref{einsteink1}).

\section{Mass definitions}
\label{appendixmass}
\subsection{Mass functions in isotropic coordinates}
Throughout the paper we used the mass function $m(R)$ defined in Eq.
 \eqref{harmonicmasseq}. In the literature it is sometimes defined another
mass function $M(R)$. The connection between the two definitions
is given below.
Using Eq. \eqref{harmonicmasseq} one gets
\beq
U(R)= 1 -G_d\int ^R\frac{m(R)} {R^{d-2}}\,dR, \label{massfunction1}
\eeq
where an integration constant has been made equal to unity.
Eq. \eqref{massfunction1} is consistent with the usual form of the
potential $U$ outside the mass and charge distributions,
i.e., $R>R_0$. In fact,
 if we take $m(R) = M=$ constant, Eq. \eqref{massfunction1} yields
$U (R)= 1 + G_dM/((d-3)R^{d-3})$, where $M$ is total mass of the source.
The other mass function of a charged dust distribution $M(R)$ can then
be defined in analogy with the result for vacuum. This is done by taking
$U(R)$
inside the dust in the same form as outside,
\beq U(R) = 1 +
\frac{G_d}{d-3}\,\frac{M(R)}{R^{d-3}}\, .
\eeq
Hence, it follows the relation
\beq
M(R) = -
(d-3)R^{d-3}\int^R \frac{m(R)}{R^{d-2}}\,dR\,. \label{massfunction2}
\eeq
And so, one sees that the two masses $m(R)$ and
$M(R)$ are in general different from each other. The two definitions
agree just in the region outside the dust fluid, in which case $m(R) =
M(R)= M$ is the total mass of the source.

\subsection{Mass functions in Schwarzschild coordinates}
The mass definition in Schwarzschild coordinates used in the paper is
given by Eq. \eqref{schwarzmasseq}. Besides such a definition, there
is a different route to define another mass function $M(r)$.  Usually,
in the literature the mass within a certain sphere of radius $r$,
$M(r)$, is defined through a relation of the form
\beq
A = \frac{1}
{1- \frac{G_d}{d-3}\frac{M(r)}{r^{d-3}}}\, .\label{massfunction3}
\eeq
Interestingly, this mass coincides with ${\cal M}(r)$ as
defined in Eq. \eqref{schwarzmasseq}.
This can be shown as follows. From the last equation it follows
\beq
M(r)= \frac{d-3}{G_d}\,r^{d-3} \left( 1 -\frac 1 A\right) \, .
\label{massfunction4}
\eeq
 Moreover, using the expression for ${dB}/{dr}$ in
terms
of $A$ obtained from \eqref{AtoBrelation} one gets
\beq
(d-3)r^{d-3} \left( 1 -\frac 1 A\right)=
 \frac{r^{d-2}}{AB}\frac{dB}{dr}\, ,\label{eqaux1}
\eeq
Therefore, comparing Eqs. \eqref{massfunction4} and \eqref{eqaux1}
one obtains
\beq
 \frac{r^{d-2}}{AB}\frac{dB}{dr}= G_d \,M(r) \, .
\eeq
Substituting
this result into Eq. \eqref{schwarzpoisson} and integrating one has
\beq
 M(r)  = S_{d-2}\,\int_0^r\rho_{m}(r)A(r)r^{d-2}dr \, ,
\eeq
which is exactly ${\cal M}(r)$ as defined in Eq. \eqref{schwarzmasseq}.
 So, one has the identity $M(r) = {\cal M}(r)$.

\end{appendix}


\end{document}